\documentclass[fleqn,usenatbib]{mnras}
\usepackage{newtxtext,newtxmath}

\usepackage[T1]{fontenc}

\DeclareRobustCommand{\VAN}[3]{#2}
\let\VANthebibliography\thebibliography
\def\thebibliography{\DeclareRobustCommand{\VAN}[3]{##3}\VANthebibliography}

\usepackage{graphicx}	
\usepackage{amsmath}	
\usepackage{acronym}
\usepackage{lineno}
\usepackage{comment}
\usepackage{xcolor}
\usepackage{tablefootnote}
\acrodef{PAUS}{Physics of the Accelerating Universe Survey}
\acrodef{NB}{narrow bands}
\acrodef{BB}{broad bands}

\title[The PAU Survey: Photometric redshift estimation in deep wide fields]{The PAU Survey: Photometric redshift estimation in deep wide fields}
\author[D.~Navarro-Gironés]{D.~Navarro-Gironés,$^{1, 2}$\thanks{E-mail: david.navarro.girones@gmail.com}
E.~Gaztañaga,$^{3,1,2}$
M.~Crocce,$^{1, 2}$
A.~Wittje,$^{4}$
H.~Hildebrandt,$^{4}$
A.~H.~Wright,$^{4}$
\newauthor
M.~Siudek,$^{1}$
M.~Eriksen,$^{5, 6}$
S.~Serrano,$^{7, 1, 2}$
P.~Renard,$^{8}$
E.~J.~Gonzalez,$^{5, 9, 10}$
C.~M.~Baugh,$^{11}$
L.~Cabayol,$^{5, 6}$
\newauthor
J.~Carretero,$^{5, 6}$
R.~Casas,$^{1, 2}$
F.~J.~Castander,$^{1, 2}$
J.~De~Vicente,$^{12}$
E.~Fernandez,$^{5}$
J.~Garc\'ia-Bellido,$^{13}$
\newauthor
H.~Hoekstra,$^{14}$
G.~Manzoni,$^{15}$
R.~Miquel,$^{5, 16}$
C.~Padilla,$^{5}$
E.~Sánchez,$^{12}$
I.~Sevilla-Noarbe,$^{12}$
\newauthor
P.~Tallada-Cresp\'{i}$^{12, 6}$
\\
$^{1}$Institute of Space Sciences (ICE, CSIC), Campus UAB, Carrer de Can Magrans, s/n, 08193 Barcelona, Spain\\
$^{2}$Institut d’Estudis Espacials de Catalunya (IEEC), Gran Capità, 2-4, Edifici Nexus, Desp. 201, 08034 Barcelona, Spain\\
$^{3}$Institute of Cosmology \& Gravitation, University of Portsmouth, Dennis Sciama Building, Burnaby Road, Portsmouth PO1 3FX, UK\\
$^{4}$Ruhr University Bochum, Faculty of Physics and Astronomy, Astronomical Institute (AIRUB), German Centre for Cosmological Lensing, 44780 Bochum, \\
Germany. \\
$^{5}$Institut de F\'isica d’Altes Energies (IFAE), The Barcelona Institute of Science and Technology, Campus UAB, 08193 Bellaterra (Barcelona), Spain.\\
$^{6}$Port d’Informaci\'o Cient\'ifica (PIC), Campus UAB, C. Albareda s/n, 08193 Bellaterra (Barcelona), Spain.\\
$^{7}$Satlantis, University Science Park, Sede Bld 48940, Leioa-Bilbao, Spain. \\
$^{8}$Department of Astronomy, Tsinghua University, Beijing 100084, China\\
$^{9}$Instituto de Astronomía Teórica y Experimental (IATE-CONICET), Laprida 854, X5000BGR, C\'ordoba, Argentina.\\
$^{10}$ Observatorio Astron\'omico de C\'ordoba, Universidad Nacional de C\'ordoba (OAC-UNC), Laprida 854, X5000BGR, C\'ordoba, Argentina.\\
$^{11}$Institute for Computational Cosmology, Department of Physics, Science Laboratories, Durham University, South Road, Durham DH1 3LE, UK.\\
$^{12}$ Centro de Investigaciones Energéticas, Medioambientales y Tecnológicas (CIEMAT), Avenida Complutense 40, 28040 Madrid, Spain.\\
$^{13}$ Instituto de Fisica Teorica UAM/CSIC, Nicolas Cabrera, 13, 28049 Madrid, Spain.\\
$^{14}$ Leiden Observatory, Leiden University, Leiden, The Netherlands.\\
$^{15}$ Jockey Club Institute for Advanced Study, The Hong Kong University of Science and Technology, Hong Kong S.A.R., China.\\
$^{16}$ Instituci\'o Catalana de Recerca i Estudis Avan\c{c}ats (ICREA), 08010 Barcelona, Spain.\\
}
\date{Accepted XXX. Received YYY; in original form ZZZ}
\pubyear{2015}
\begin{document}
\label{firstpage}
\pagerange{\pageref{firstpage}--\pageref{lastpage}}
\maketitle

\begin{abstract}
We present photometric redshifts (photo-$z$) for the deep wide fields of the Physics of the Accelerating Universe Survey (PAUS), covering an area of $\sim$50 deg$^{2}$, for $\sim$1.8 million objects up to $i_{\textrm{AB}}<23$. The PAUS deep wide fields overlap with the W1 and W3 fields from CFHTLenS and the G09 field from KiDS/GAMA. Photo-$z$ are estimated using the 40 narrow bands (NB) of PAUS and the broad bands (BB) of CFHTLenS and KiDS. We compute the redshifts with the SED template-fitting code \textsc{BCNZ}, with a modification in the calibration technique of the zero-point between the observed and the modelled fluxes, that removes any dependence on spectroscopic redshift samples. We enhance the redshift accuracy by introducing an additional photo-$z$ estimate ($z_{\textrm{b}}$), obtained through the combination of the \textsc{BCNZ} and the BB-only photo-$z$. Comparing with spectroscopic redshifts estimates ($z_{\textrm{s}}$), we obtain a $\sigma_{68} \simeq 0.019$ for all galaxies with $i_{\textrm{AB}}<23$ and a typical bias $|z_{\textrm{b}}-z_{\textrm{s}}|$ smaller than 0.01. For $z_{\textrm{b}} \sim (0.10-0.75)$ we find $\sigma_{68} \simeq (0.003-0.02)$, this is a factor of $10-2$ higher accuracy than the corresponding BB-only results. We obtain similar performance when we split the samples into red (passive) and blue (active) galaxies. We validate the redshift probability $p(z)$ obtained by \textsc{BCNZ} and compare its performance with that of $z_{\textrm{b}}$. These photo-$z$ catalogues will facilitate important science cases, such as the study of galaxy clustering and intrinsic alignment at high redshifts ($z \lesssim 1$) and faint magnitudes.

\end{abstract}



\section{Introduction}

Over the last decades, wide-field galaxy surveys, such as the Sloan Digital Sky Survey (SDSS, \citealt{SDSS}), the Canada-France-Hawaii Telescope Legacy Survey (CFHTLS, \citealt{CFHTLs}), the Kilo-Degree Survey (KiDS, \citealt{KiDS}), the Dark Energy Survey (DES, \citealt{DES_2016}) and the Hyper Suprime-Cam Subaru Strategic Program (HSC SSP, \citealt{HSC}), amongst others, have provided the community with a large number of photometric redshift measurements, which allowed to perform a range of cosmological statistical analyses, such as estimating correlation functions for galaxy clustering or cosmic shear, which might help to infer the nature of dark energy and dark matter (\citealt{Weinberg}).

Redshifts can be estimated using spectroscopic or photometric techniques. The spectra of an object can be well determined by the former yielding a precise redshift determination, mainly through the detection of emission and absorption lines, the Lyman break or the 4000\r{A} break. However, this technique is expensive, in the sense that measurements have to be made per object (or a limited number of objects for recent spectroscopic surveys such as \citealt{DESI}), knowing its angular coordinates beforehand and with large cost in exposure times due to the inefficiencies of spectroscopy. On the other hand, photometric surveys measure the flux of many objects at once, using filters that allow us to recover a low resolution spectrum of each object. The resolution at which the spectrum is obtained is limited by the wavelength width of the filter, so that narrower bands have greater resolution. Typically, photometric surveys use optical bands with a width of $\sim$100nm, we will refer to those as broad band (BB) photometric surveys.

Here, we take advantage of photometry from the Physics of the Accelerating Universe Survey (PAUS, \citealt{Padilla2019}), which is composed of 40 narrow bands (NB) with a width of $\sim$13nm. With bands one order of magnitude narrower than that of broad bands, the spectra recovered by PAUS have a resolution between spectroscopic and broad band photometric surveys, which allows us to compute photometric redshifts (photo-$z$) with unprecedented precision. Other narrow band photometric surveys include ALHAMBRA (\citealt{Alhambra}), mini-JPAS (\citealt{miniJPAS}) and LAGER (\citealt{LAGER}).

The main techniques used to estimate photometric redshifts are spectral energy distribution (SED) template-fitting codes (\citealt{Benitez2000, Hyperz, Zebra, LePhare, CIGALE, Eriksen2019}) and machine learning algorithms (\citealt{ANNz, SOM_machine_learning, TPZ, NeuralNetworks_machine_learning, SupportVector_machine_learning, Eriksen_machine_learning}). The former method compares the measured fluxes with a set of SED templates at different redshifts and for different kinds of galaxy populations, including emission lines as well as the stellar continuum. In this case, a library of SED templates is needed to represent the variety of galaxy populations, with the performance of the method relying in the completeness of these SED templates. However, some colour-redshift degeneracies might appear that will affect the performance of such codes. In the latter case, machine learning algorithms are trained with fluxes of galaxies for which we know their spectroscopic redshifts. The performance of this method is limited to the goodness of the training data, hence the spectroscopic sample needs to realistically represent the photometric one. Here, we will use a SED template-fitting code called \textsc{BCNz2} (\citealt{Eriksen2019}), which we will refer to as \textsc{BCNZ} to simplify the notation throughout the paper,  specifically designed to deal with the 40 NB of PAUS.

Previous studies of photometric redshifts in PAUS have been performed only in the COSMOS field with SED template-fitting codes (\citealt{Eriksen2019, Alarcon_bcnz}) and machine learning algorithms (\citealt{Eriksen_machine_learning, John_Soo, Laura_Cabayol}), covering an area of $\sim$1.5 deg$^{2}$, down to $i_{\textrm{AB}}<22.5$ in the case of \citet{Eriksen2019}, \citet{John_Soo} and \citet{Eriksen_machine_learning} and $i_{\textrm{AB}}<23$ in the case of \citet{Alarcon_bcnz} and \citet{Laura_Cabayol}. Here, we aim to release photometric redshift results for $\sim$1.8 million objects in 51 deg$^{2}$ down to $i_{\textrm{AB}} < $ 23, with redshifts from 0 to 2.

The photometric redshifts we present here have already been used for scientific purposes, such as the study of the D4000 spectral break index in PAUS (\citealt{D4000}), the identification of close galaxy pairs and the determination of their mean mass (\citealt{GalaxyPairs}) and the evaluation of the capability prediction of semi-analytical galaxy formation models using the spectroscopic samples presented in this work (\citealt{Manzoni}). In Wittje et al. (in prep.) they simulate the PAUS fluxes using the Flagship simulation (\citealt{Flagship}) and include a comparison between the performance of the photometric redshifts computed from the simulation and the estimates presented in this work. Here, we aim to describe the process of obtaining the new photo-$z$ for the deep wide fields of PAUS and study their performance as a function of magnitude, redshift, colour and photometric quality. Relevant projects will be performed thanks to the PAUS data presented in this paper, such as the study of the intrinsic alignments of galaxies (D. Navarro-Gironés et al., in preparation) or the study of galaxy clustering through density maps (Gonzalez et al., in preparation), amongst others.

The structure of this paper is as follows. In Section~\ref{sec:Data}, we present the data we will use to obtain the fluxes of the objects and the spectroscopic information to validate the photometric redshifts. In Section~\ref{sec:Methodology}, we will explain the methodology used to obtain the photo-$z$, explaining the SED template-fitting code we have used and some of the improvements and adaptations that we implemented to it. In Section~\ref{sec:Photo-z catalogues}, we will present and validate the photo-$z$, we will study its performance as a function of the galaxy colours and we will analyze the $p(z)$ distributions. Finally, in Section~\ref{sec:Conclusions} we will close with some conclusions.

\section{Data}
\label{sec:Data}

\subsection{PAUS}
\label{sec:PAUS}

\begin{figure}
    \centering
    \includegraphics[width=0.45\textwidth]{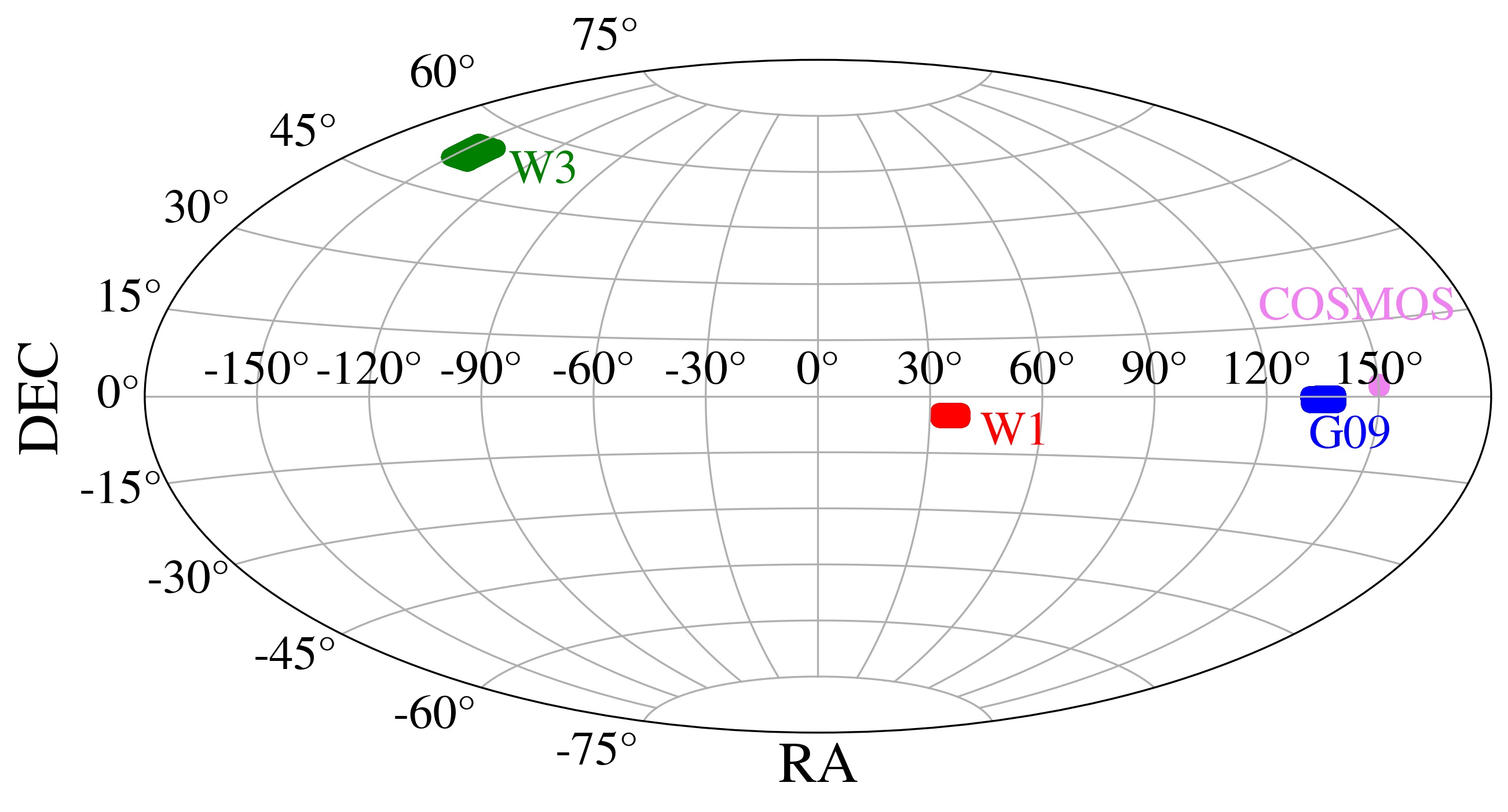}
    \caption{Position in the sky of the W1, W3, G09 and KiDZ-COSMOS fields used in this study.}
    \label{fig:Area}
\end{figure}

PAUS is a photometric survey conducted at the William Herschel Telescope at El Roque de Los Muchachos, in the Canary Islands. The PAUCamera (\citealt{Padilla2019}) used a set of 40 NB filters ranging from 4500\r{A} to 8500\r{A} in steps of 100\r{A} with a FWHM of 130\r{A}. This exceptional filter arrangement is designed to provide high-precision photo-$z$, outperforming the precision achievable with broad band observations. PAUS targets are comprised of the COSMOS field (\citealt{COSMOS_field}), which is mainly used for calibration and validation processes, the W1, W3 and W4 Wide Fields from the Canada-France-Hawaii Telescope Lensing Survey \citep[CFHTLenS;][]{CFHTLenS, CFHTLenS_Heymans} and the GAMA G09 field, which overlaps with the Kilo-Degree Survey \citep[KiDS;][]{KiDS}. The overlap between the W1, W3 and G09 fields with PAUS observations compose the PAUS deep wide fields, which we will refer to as PAUS wide fields for brevity. Fig.~\ref{fig:Area} shows the position in the sky of the PAUS targets. In our study, we exclude the W4 field since the observations made by PAUS in it are rather scarce. PAUS covers an area of $\sim$43 deg$^{2}$ in all 40 NB and $\sim$51 deg$^{2}$ with a coverage of at least 30 NB, up to $i_{\textrm{AB}} = 23$ and with $\sim$1.8 million objects observed (see Table \ref{tab:PAUS_target}). The number density of PAUS objects with accurately measured photo-$z$ is $\simeq 3 \times 10^4$ per square degree. This number is large compared to 
current wide spectroscopic surveys, such as GAMA (\citealt{GAMA}) or VIPERS (\citealt{VIPERS}), with number densities $\sim$10$^3$, which are not as deep as PAUS and usually not complete in magnitude. Although spectroscopic surveys can have more precise redshifts and larger area coverage, PAUS regime is useful when we need larger densities over wide fields, with photo-$z$ that are more accurate than the ones obtained from broad bands. In combination with the 40 NB, PAUS  observations are complemented with BB photometry provided by CFHTLenS and KiDS, which are also used as reference samples because they are deeper than PAUS (see Section~\ref{sec:BBphot}).

PAUS fluxes are measured by doing forced photometry over these reference samples. This technique takes the positions and galaxy shapes in the reference samples
and measures the NB fluxes at these fixed positions. The objects in the PAUS wide fields are observed an average of 3 times. These observations are later coadded at the catalogue level, obtaining the flux and its error per object, which will be used for science. For a detailed explanation of this procedure and the PAUS flux calibration, we refer the reader to \cite{Serrano2022} and Castander et al. (submitted to MNRAS).

\begin{table}
\caption{Area with a minimum coverage of 30 NB, number of objects up to $i_{\textrm{AB}}<23$, RA and DEC limits of the PAUS targets.}
\begin{center}
\label{tab:PAUS_target}
\resizebox{\columnwidth}{!}{\begin{tabular}{c c c c c}
\hline
\hline
Field & Area 30 NB & \# objects & RA limits     & DEC limits       \\
 & (deg$^{2}$) &  & (deg) & (deg)\\ \hline
W1    & 12.04    & 401815     & {[}32.3,38.5{]} & {[}-6.1, -4{]}   \\ \hline
G09   & 15.7    & 663535     & {[}131,139{]}   & {[}-1.7, 0.5{]}  \\ \hline
W3    & 22.64    & 792664     & {[}209,219.5{]} & {[}51.4, 55.6{]} \\ \hline
KiDZ-COSMOS \tablefootnote{KiDZ-COSMOS refer to objects from the COSMOS field with spectroscopic information, as will be explained in Section~\ref{sec:Spectroscopic}}    & 1    & 13380     & {[}149.5,150.6{]} & {[}1.7, 2.7{]} \\ \hline
Total    &  51.38   &   1871394   & - & - \\ \hline
\end{tabular}}
\end{center}
\end{table}

Finally, the treatment of the data is carried out by the PAU data management (PAUdm) team, located at Port d’Informació Científica (PIC). The responsibilities of PAUdm include the storage, data reduction and accessibility of PAUS measurements to its members. The PAU database, where the data and metadata of PAUS measurements and results are stored, is of utmost importance. For a detailed description of the design and responsibilities of PAUdm we refer the reader to \cite{PAUdm}.

\subsection{Broad Band photometry}
\label{sec:BBphot}

As explained in Section~\ref{sec:PAUS}, broad band photometry is needed to define the positions and detect the sources in PAUS and also to add extra information coming from the BB fluxes. These BB photometric catalogues are also called parent catalogues or reference catalogues, given that they define the samples to be observed. The Wide Fields from CFHTLenS and the GAMA G09 field from KiDS constitute the parent catalogues used by PAUS.

The CFHTLS-Wide uses the wide-field imager, MEGACAM \citep{CFHTLs_cam}, installed at Mauna Kea. Its field of view covers 1 deg$^{2}$ in the \textit{ugriz} (\textit{y} after the \textit{i} filter broke) broad band filters up to a $5\sigma$ limiting magnitude of $i_{\rm AB} \approx$ 25.5. The CFHTLS-Wide observes in the W1, W2, W3 and W4 fields, covering 157 deg$^{2}$. In particular, the CFHTLenS team was formed to conduct weak-lensing studies and the measuring of the galaxy shapes was implemented by the CFHTLenS shape measurement pipeline (\citealt{CFHTLenS_shapes}). The best observing conditions were reserved for the \textit{i}-band, making it the survey detection band. CFHTLenS multi-band photometry was extracted from PSF-homogenised \citep[][]{2008A&A...482.1053K}{}{} stacks with SExtractor \citep[][]{1996A&AS..117..393B}{}{} in dual-image mode yielding high-quality colours for photo-$z$ estimates (\citealt{CFHTLs_photoz}).

KiDS is a wide-field imaging survey that uses OmegaCam (\citealt{OmegaCam}), which is installed at the ESO VLT Survey Telescope (\citealt{VLT}) at the Paranal observatory. OmegaCam has a field of view of 1 deg$^{2}$ in the \textit{ugri} bands, with the \textit{r}-band being the one used in best conditions to enable the precise measurement of galaxy shapes. Matching objects between the fields observed by KiDS and those observed by the VISTA Kilo-degree INfrared Galaxy (VIKING) survey (\citealt{VIKING}), allows the addition of VISTA's five near-infrared broad bands \textit{ZYJHK$_{\rm s}$}. KiDS DR4 covers around 1000 deg$^{2}$ in the KiDS fields, KiDS-S and KiDS-N. In particular, KiDS-N overlaps with the G09 field of GAMA (\citealt{GAMA}), which we will use in our study. The KiDS multi-band photometry is extracted with the GAaP method \citep[][]{2008A&A...482.1053K}{}{}, which first convolves each image with a shapelet-based kernel to yield a Gaussian PSF and then measures fluxes in Gaussian-weighted elliptical apertures (\citealt{KiDS_DR4}).

\begin{figure}
    \centering
    \includegraphics[width=0.48\textwidth]{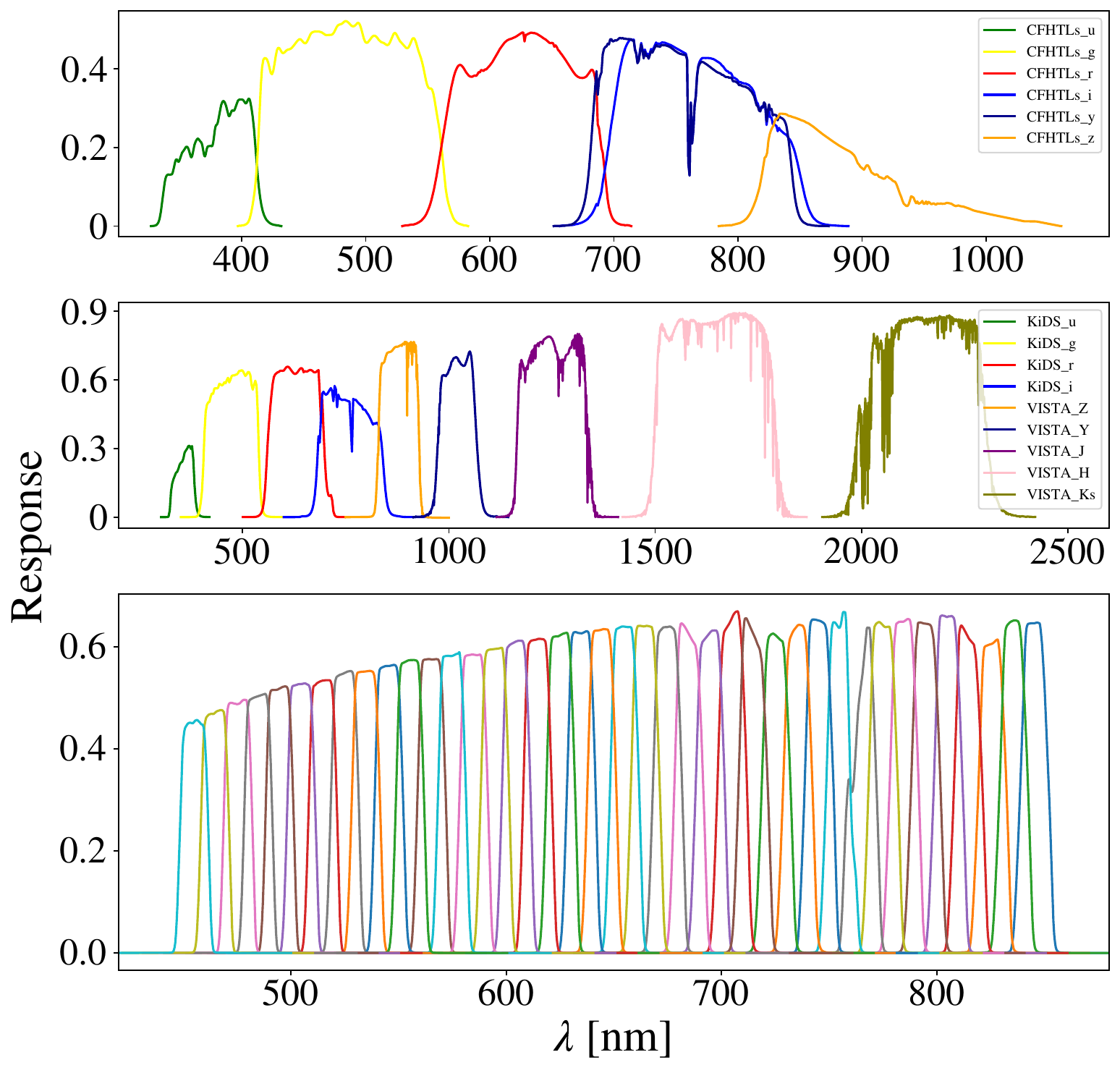}
    \caption{Response of the filters used in this study as a function of the wavelength (nm) for the CFHTLenS (top) and the KiDS (middle) broad bands and for the PAUS narrow bands (bottom).}
    \label{fig:Filters}
\end{figure}

Fig.~\ref{fig:Filters} shows the filter response as a function of the wavelength of the NB used by PAUS (lower panel) and the BB used by both CFHTLenS (top panel) and KiDS (middle panel). The wide coverage in wavelength of both broad and narrow bands used in this study is one key component that allows highly accurate determination of the photometric redshifts. The narrow band wavelength range ($450-850 $nm) overlaps with the $g$, $r$, $i$, $y$ and, partially, $z$ CFHTLenS BB and by the $g$, $r$, $i$ and $Z$ KiDS BB, while the rest of the BB widen the wavelength coverage of PAUS narrow bands. 

\begin{figure}
    \centering
    \includegraphics[width=0.48\textwidth]{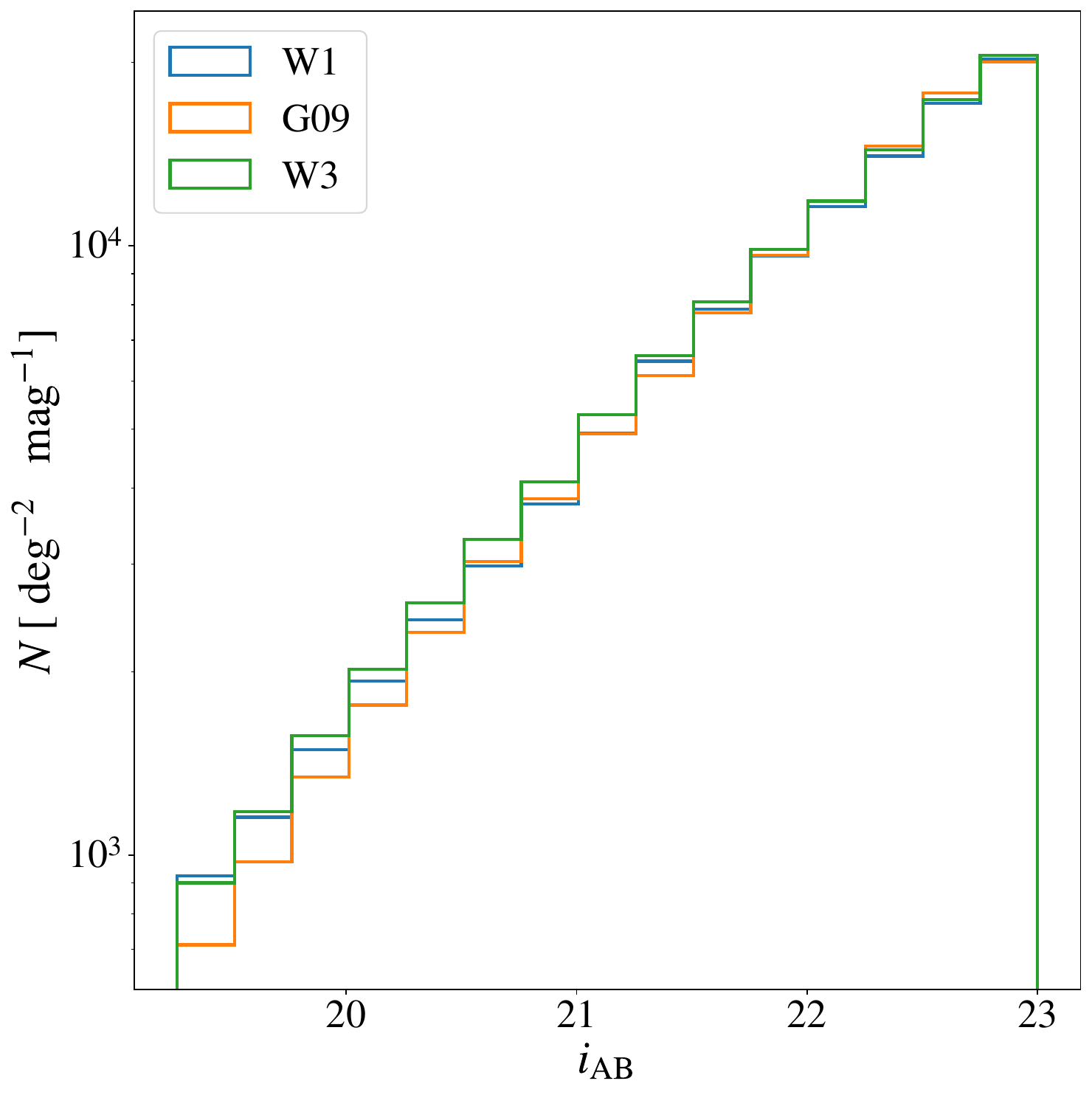}
    \caption{Number counts in the three PAUS wide fields as a function of $i_{\textrm{AB}}$, where $i_{\textrm{AB}} \equiv i_{\textrm{AB, CFHTLenS}} = i_{\textrm{AB, KiDS}} - 0.1$. The fact that the number counts for the G09 field are very similar to those for the W1 and W3 fields over the considered magnitude range, indicates that the limit imposed at $i_{\textrm{AB, KiDS}} = 23.1$ in the selection is appropriate.}
    \label{fig:number_counts}
\end{figure}

The CFHTLenS $i$-band filter broke and was replaced by a similar filter that was labeled as the $y$-band. For objects that were not measured with the $i$-band, we will use $y$-band measurements and treat them equally as the $i$-band measurements (see how similar the transmission curves are for the $i$ and $y$ bands in the top panel of Fig. \ref{fig:Filters}). Another important aspect to take into account is that the filter response functions and the wavelength range covered by the $i$-bands used in CFHTLenS and KiDS are not the same. As a result, the $i_{\textrm{AB}}$ of both systems is defined slightly differently. In order to select a similar population for all fields, we need to find a relation between $i_{\textrm{AB, KiDS}}$ and $i_{\textrm{AB, CFHTLenS}}$ (the $i_{\textrm{AB}}$ magnitude in the KiDS and the CFHTLenS systems, respectively). This relation is established by studying the number counts ($N$) of the PAUS wide fields and determining what is the selection cut to be applied in $i_{\textrm{AB, KiDS}}$ that corresponds to the number count at $i_{\textrm{AB, CFHTLenS}}=23$. Fig.~\ref{fig:number_counts} shows the number counts for the three PAUS wide fields as a function of the magnitude $i_{\textrm{AB, CFHTLenS}}$, which from now on we will refer to as $i_{\textrm{AB}}$, if not specified otherwise. We find that applying a cut at $i_{\textrm{AB, KiDS}} = 23.1$ and redefining $i_{\textrm{AB}} \equiv i_{\textrm{AB, CFHTLenS}} = i_{\textrm{AB, KiDS}} -0.1$ gives reasonably similar number count values between the G09 and the W1 and W3 fields over the whole magnitude range. However, there are still some differences between the G09 counts and those in the W1 and W3 fields that may be due to the different definitions used for the $i$-band (given that the area of the fields is relatively small this could also be due to sample variance). This indicates that a comparison of the photo-$z$ performance between fields as a function of $i_{\textrm{AB}}$ is not straightforward, as we can see in Appendix~\ref{sec:Study fields}.

\begin{figure*}
    \centering
    \includegraphics[width=.8\textwidth]{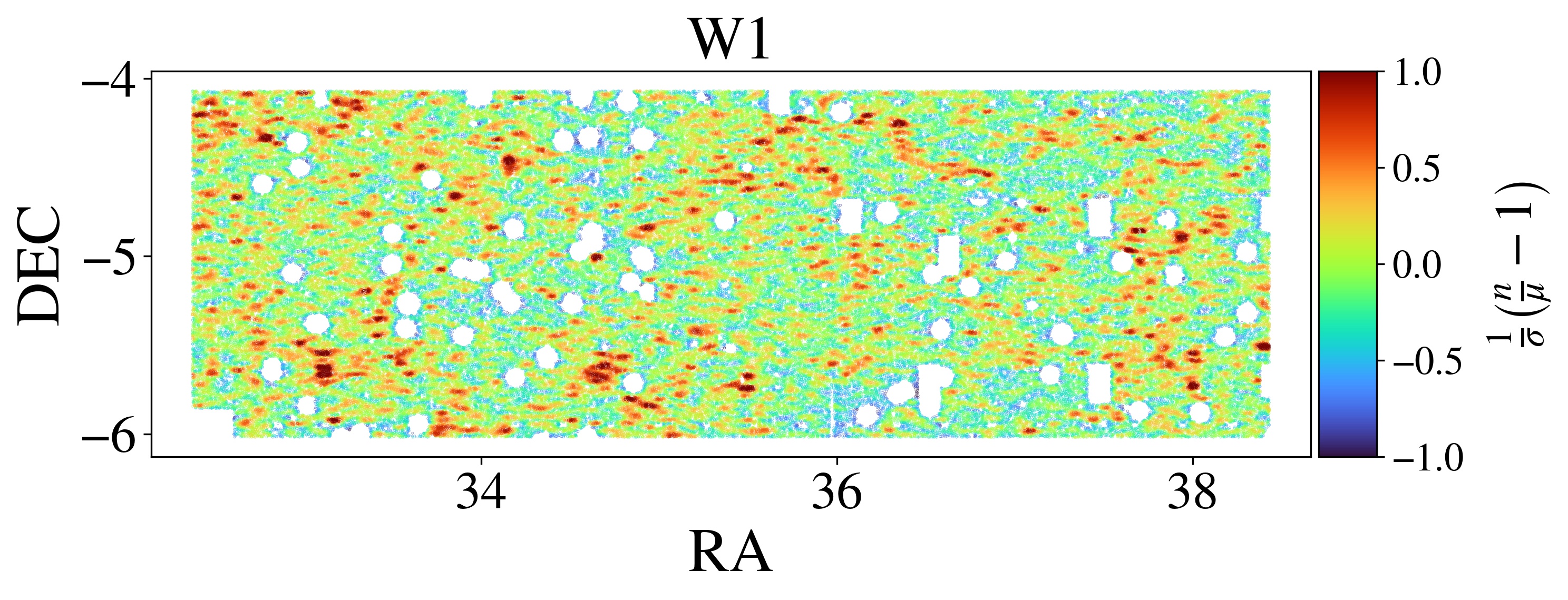}
    \includegraphics[width=\textwidth]{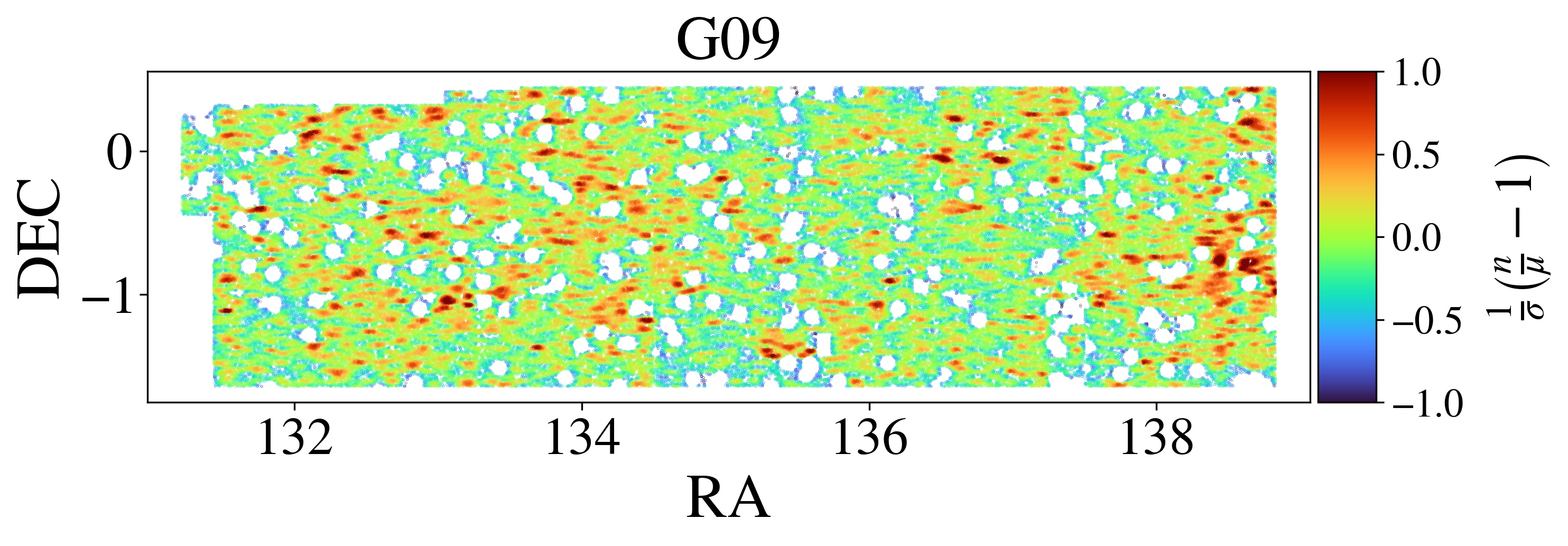}
    \includegraphics[width=\textwidth]{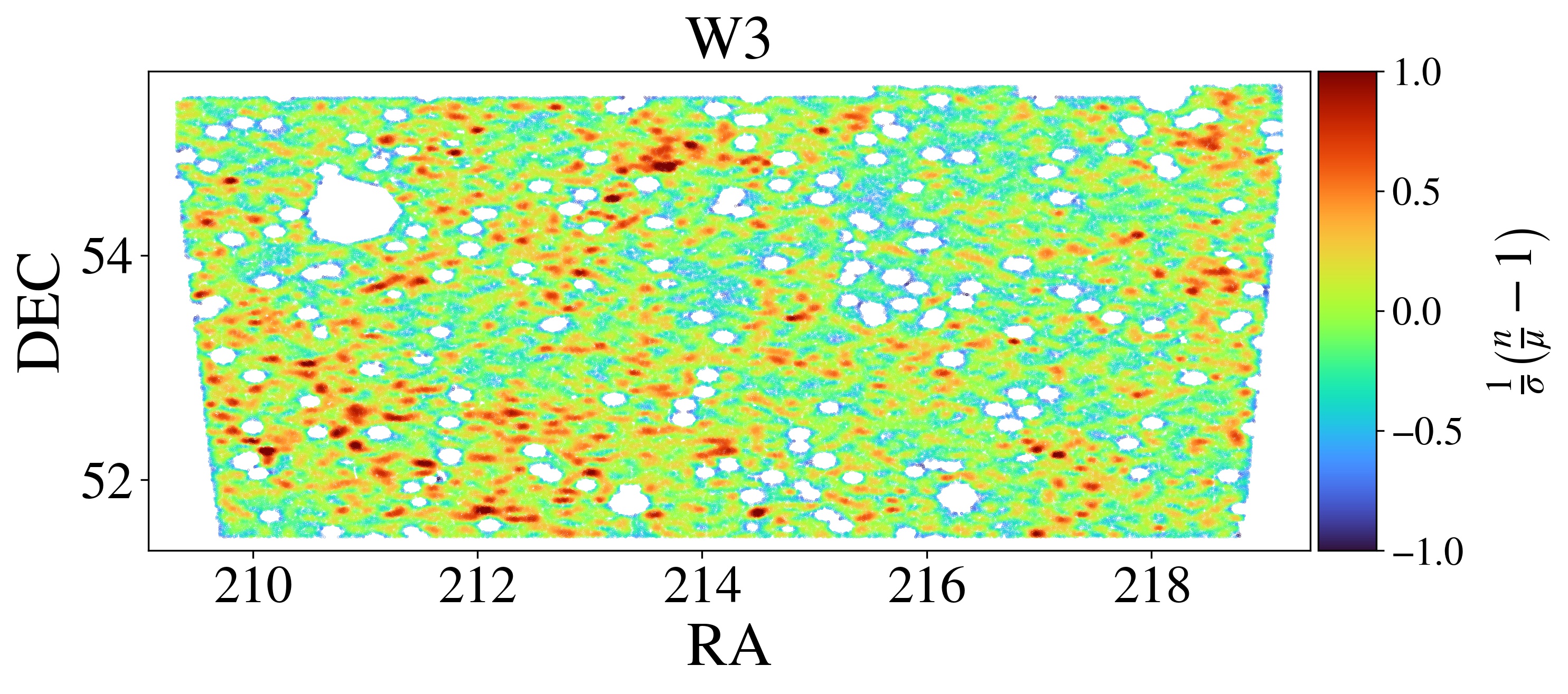}
    \caption{Angular distribution of galaxies in the W1, G09 and W3 fields after applying the reference catalogue star flags and masks, which remove stars and objects not well identified or with bad photometry. The colour code indicates the fluctuations in the number density of objects ($n$), $\frac{1}{\sigma}\left ( \frac{n}{\mu}-1\right )$, such that $\sigma$ and $\mu$ correspond, respectively, to the $\sigma_{68}$ and the median value of $n$.}
    \label{fig:Area_density}
\end{figure*}

To perform the analysis, we remove from the reference catalogues the objects with bad quality photometry or those classified as stars. In the case of CFHTLenS, we exclude stars by setting the star\_flag $=$ 0 and the mask from CFHTLenS less or equal than 1. In the case of KiDS, we set sg\_flag $=$ 0 and sg2dphot $=$ 1 to remove stars. We also apply a series of masks related to the detection band $r$, which perform a stellar masking, mask due to saturation, trim and account for chip gaps, void mask and asteroids. Fig.~\ref{fig:Area_density} shows the distribution of galaxies as a function of RA and DEC in the W1, W3 and G09 fields coloured by the fluctuations in the number density of objects (defined as the ratio between the number of objects and the area), after applying the reference catalogue masks and star flags. Here the fluctuations are computed as $\frac{1}{\sigma}\left ( \frac{n}{\mu}-1\right )$, with $\sigma$ and $\mu$ corresponding, respectively, to the $\sigma_{68}$ and the median value of the number density, $n$. It is interesting to observe the angular clustering of galaxies through the overdensities (red regions) and voids (bluer regions).

We also apply an extinction correction to the fluxes ($\phi$) that takes into account the Milky Way extinction:
\begin{equation}\label{eq:GalacticExtinction}
    \phi_{corr} = \phi_{uncorr} \cdot \frac{1}{C_{0} \cdot E(B-V)^{2} + C_{1} \cdot E(B-V) + C_{2}},
\end{equation}
where the $E(B-V)$ values are extracted from the \cite{PlanckCollaboration2013} dust map and $C_{0}$, $C_{1}$ and $C_{2}$ are band dependent extinction coefficients.

The fluxes from the reference catalogues were already extinction corrected by the CFHTLenS and KiDS teams, while the PAUS fluxes were not. To apply the same extinction method in all bands, we first added the extinction that was already corrected for in the broad bands and later corrected all NB and BB using the Milky Way extinction model defined in eq.~\ref{eq:GalacticExtinction}.

\subsection{Spectroscopic data}
\label{sec:Spectroscopic}

Spectroscopic redshifts (spec-$z$) are essential for validating the photo-$z$ performance. This validation involves comparing the spectroscopic redshifts, which have greater accuracy, with their photometric counterparts. We refer to the subsamples with spec-$z$ information as validation samples, which are defined after applying the survey mask and the stellar flags mentioned in Section~\ref{sec:BBphot}. Due to the wide angular separation of the PAUS wide fields, we need spectroscopic data from different galaxy surveys. The properties of the main spectroscopic redshift surveys included in this analysis are described in the following lines:
\begin{enumerate}
\item The Sloan Digital Sky Survey (SDSS) is a wide-area spectroscopic survey conducted at the Apache Point Observatory (APO) and Las Campanas Observatory (LCO). We use data from the DR16 (\citealt{SDSS}), with redshifts ranging from 0 to 1.1 and mag-$i$ limit $\sim$22. We select galaxies based on the flag CLASS == 'GALAXY' and zWarning == 0, indicating that there are not unknown associated problems.
\item The Galaxy and Mass Assembly (GAMA) (\citealt{GAMA_general}) spectroscopic survey observed galaxies over $\sim$286 deg$^{2}$ with a flux limit of mag-$r$ $=19.8$ and a redshift distribution that extends to $z=0.5$. We use the DR3 (\citealt{GAMA_DR3}), which mainly covers until redshift 0.5 and we select the best spectroscopic redshifts by setting the quality parameter \textit{nQ} $>= 3$.
\item The VIMOS Public Extragalactic Redshift Survey (VIPERS) (\citealt{VIPERS}) was  performed at ESO's Very Large Telescope (VLT) in Chile. The survey magnitude limit is $i_{\textrm{ab}}=22.5$ and covers a redshift range of $0.5<z<1.2$ (as targets are colour selected to lie in this range) over an area of $\sim$23.5 deg$^2$. The redshift quality flag $3 \leq \textrm{zflg} \leq 4$ is applied.
\item The DEEP2 redshift survey (\citealt{DEEP2_1}, \citealt{DEEP2_2}) used the DEIMOS spectrograph at the Keck-II telescope. This survey covers $\sim$2.8 deg$^2$ in four fields and observes objects out to $z \sim 1$ up to a limiting magnitude $R_\textrm{AB} = 24.1$. The quality flag $3 \leq \textrm{zquality} \leq 4$ is applied.
\item KiDZ-COSMOS objects are extracted from the KiDS DR5 (Wright et al., in press) spectroscopic sample in the COSMOS field. They are mainly provided from G10-COSMOS (\citealt{G10_COSMOS}), a re-reduction of the zCOSMOS-bright sample (\citealt{COSMOS2007, COSMOS2009}), which covers 1.7 deg$^{2}$ up to $i_{\textrm{AB}}<22.5$ and $0.1<z<1.2$. All objects have high quality spectroscopic redshifts with \textit{nQ} $>= 3$.
\item The 2dFGRS (Two-degree Field Galaxy Redshift Survey) (\citealt{2dfGRS}) was observed from the Anglo-Australian Observatory and used the 2dF spectrograph (\citealt{2df}), which covers a 2 degree diameter field of view. It measured $\sim$250000 galaxies, covering an area of 2000 deg$^{2}$ with a limiting magnitude of $b_{J} \sim 19.45$ and a median redshift of $z = 0.11$.
\item The VIMOS VLT DEEP Survey (VVDS) (\citealt{VVDS}) is a magnitude limit spectroscopic redshift survey that has observed 34594 objects with spec-$z$ from $0 \leq z \leq 6.7$ up to $i_{\textrm{AB}} \sim 24.75$.
\item The 3D-HST (\citealt{3DHST}) is a spectroscopic survey with the Hubble Space Telescope specially designed to study galaxy formation at $1 \leq z \leq 3.5$. It presents a 5$\sigma$ signal-to-noise ratio (SNR) per resolution element up to $H_{140}\sim23.1$, the F140W filter imaging.
\end{enumerate}

Table~\ref{tab:spec} shows the number of objects of the main spectroscopic surveys from which the data is taken to validate the three PAUS wide fields. The W1 field from CFHTLenS is mainly covered by VIPERS, GAMA, SDSS, VVDS, 2dFGRS and 3D-HST. The G09 field is covered by GAMA and SDSS. However, as will be seen at the end of this section, some objects from the COSMOS field, which does not cover G09, are also used to validate that field. The reason for this is that both COSMOS and G09 photo-$z$ are run within the same photometric system, as will be explained below. Finally, W3 overlaps with DEEP2, SDSS and 3D-HST. Another 1323 redshifts come from miscellaneous sources\footnote{The miscellaneous sources are mainly composed by UDSz (\citealt{UDSz1, UDSz2}), C3R2 (\citealt{C3R2}), IMACS (\citealt{IMACS}), VANDELS (\citealt{VANDELS}) and SAGA (\citealt{SAGA}).} not included in Table~\ref{tab:spec} for brevity.

\begin{table}
\caption{Main spectroscopic redshift surveys used in the W1, W3 and G09 fields. The first column gives the name of each spectroscopic survey. The second, third and fourth columns give the number of spectroscopic redshifts in each field.}
\begin{center}
\begin{tabular}{c c c c}
\hline
Survey & W1    & W3   & G09  \\ \hline
SDSS         & 5437  & 8018 & 1213 \\ 
GAMA         & 8884  & 0    & 4704 \\ 
VIPERS       & 21378 & 0    & 0    \\ 
DEEP2        & 0     & 6969 & 0    \\ 
KiDZ-COSMOS  & 0     & 0 & 11854    \\ 
2dFGRS & 2662  & 0    & 0    \\ 
VVDS         & 2216  & 0    & 0    \\ 
3DHST        & 933   & 707  & 0    \\ 
Miscellaneous        & 1193   & 130  & 0    \\ \hline
Total        & 42703 & 15824 & 17771    \\ \hline
\end{tabular}
\end{center}
\label{tab:spec}
\end{table}

\begin{figure}
    \centering
    \includegraphics[width=0.47\textwidth]{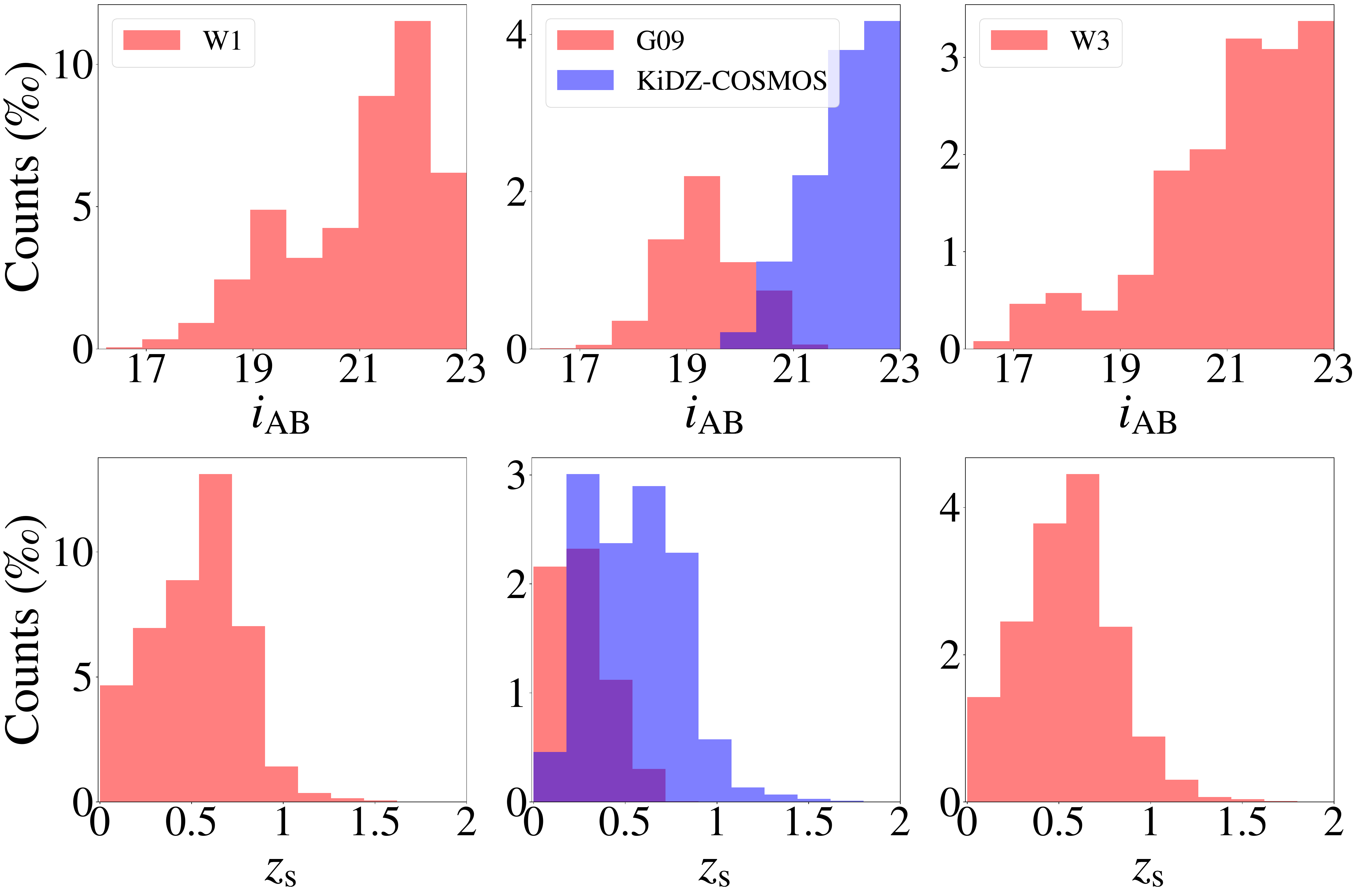}
    \caption{$i_{\textrm{AB}}$ magnitude (top) and spec-$z$ (bottom) distributions of the validation samples for W1, G09, KiDZ-COSMOS and W3 (from left to right, as labelled in the top row panels). The validation sample of the G09 field does not cover the necessary range either in $i_{\textrm{AB}}$ or spec-$z$, making it unsuitable for validating the G09 photometric redshifts by itself. The coverage in $i_{\textrm{AB}}$ and spec-$z$ for KiDZ-COSMOS complements that for the G09 field and allows it to cover deeper regions.}
    \label{fig:Validation_samples}
\end{figure}

Fig.~\ref{fig:Validation_samples} shows the distribution of $i_{\textrm{AB}}$ and spec-$z$ for the objects with spectroscopic redshifts for the three fields under study. The shape of these distributions is important when assessing the photometric redshift performance in each field as a function of $i_{\textrm{AB}}$ or spec-$z$. Low counts in a given bin may lead to poor statistics in the determination of the performance. Ideally, the distribution of the validation samples should be very similar to the whole catalogue. However, this is not always possible due to the lack of spectroscopic redshifts available. In the case of the W1 and W3 fields, the distributions look quite similar in terms of $i_{\textrm{AB}}$ and redshift, with the exception that there is a drop-off in the number of objects for the last magnitude bin ($i_{\textrm{AB}} \sim 22.5$) in the W1 field that is not present in W3, where the number of objects above $i_{\textrm{AB}}\sim 21$ remains fairly constant. In the case of the G09 validation sample, one can see that the coverage in both $i_{\textrm{AB}}$ and redshift is quite poor. 

The G09 validation sample barely goes beyond $i_{\textrm{AB}}\sim 21$ and redshift $\sim$0.75. This poses a challenge when validating the performance of the photo-$z$ in this  field. To overcome this problem, we indirectly validate the G09 field by computing new photometric redshifts in the COSMOS field, using KiDS as a reference catalogue and using its broad band photometry (we will refer to this catalogue as KiDZ-COSMOS). This validation is possible because both PAUS and KiDS observe in the COSMOS field. By doing that, we are using the same photometric system either in the G09 and KiDZ-COSMOS fields. We also compute these photo-$z$ using the same magnitude depth in both cases. This validation process will be addressed in detail in Section~\ref{sec:KIDZ_COSMOS}. The number of KiDZ-COSMOS objects (listed in Table~\ref{tab:spec}) nearly doubles the number of objects in the G09 validation sample.

Fig.~\ref{fig:Validation_samples} shows that KiDZ-COSMOS covers a range in $i_{\textrm{AB}}$ and spec-$z$ not covered before by the G09 field, allowing us to span the $i_{\textrm{AB}}$ and the spectroscopic ranges of the W1 and W3 fields. However, the combined G09 and KiDZ-COSMOS validation samples, have a pronounced peak at $i_{\textrm{AB}} \sim 19$ and the spec-$z$ distribution is different from the W1 and W3 cases. This fact may result in a more intricate comparison of the G09 field's performance in relation to the W1 and W3 fields.

It is also important to note that the coverage in spec-$z$ for all three fields decreases drastically beyond $z_{\textrm{s}} = 1$. For this reason, we will restrict our validation to  $z_{\textrm{s}}<1.5$ in Section~\ref{sec:Photo-z catalogues}. This is a challenge to validate our results, since the photo-$z$ code we use allows us to compute redshifts until $z_{\textrm{b}} = 2$. However, due to the magnitude limit at $i_{\textrm{AB}} = 23$, there will not be that many objects at $z_{\textrm{b}} > 1.5$.

\section{Methodology}
\label{sec:Methodology}

\subsection{\textsc{BCNZ}}
\label{sec:bcnz}

The estimation of the photo-$z$ is  performed using a template based code called \textsc{BCNZ} (\citealt{Eriksen2019}), in which the observed flux of a galaxy is fitted against a linear combination of redshift-dependent templates. This code has been specifically designed to process the information coming from both the NB and BB data, also incorporating emission lines. In what follows we will briefly explain how \textsc{BCNZ} computes the photometric redshifts \citep[for a detailed explanation see][]{Eriksen2019}{}{}.

For each galaxy, the probability redshift distribution is obtained via:
\begin{equation}\label{eq:pz}
    p(z) \propto \int _{\alpha_{1}\geq 0} \textrm{d}\alpha_{1} \cdot\cdot\cdot \int _{\alpha_{n}\geq 0} \textrm{d}\alpha_{n} e^{-0.5\chi^{2}(z, \alpha)} p_{\rm{prior}}(z, \alpha),
\end{equation}
where $\alpha_{i}$ corresponds to the amplitude of template $i=1,\dots,n$, $p_{\rm{prior}}$ is the form of the priors and $\chi^{2}$ is defined as:
\begin{equation}\label{eq:chi2}
    \chi^{2}(z, \alpha) = \sum_{i, \text{NB}}\left ( \frac{\phi_{i}^{\text{obs}} - l_{i}k\phi_{i}^{\text{model}}}{\phi_{\textrm{err},i}} \right )^{2} + \sum_{i, \text{BB}}\left ( \frac{\phi_{i}^{\text{obs}} - l_{i}\phi_{i}^{\text{model}}}{\phi_{\textrm{err},i}} \right )^{2},
\end{equation}

where $\phi_{i}^{\text{obs}}$ and $\phi_{i}^{\text{model}}$ are the observed and modelled fluxes in band $i$, respectively; $\phi_{\textrm{err}, i}$ is the error on  $\phi_{i}^{\text{obs}}$ and $k$ and $l$ are calibration parameters.

On the one hand, $k$ is intended to calibrate between the NB and BB fluxes for each galaxy and can be easily obtained by taking the derivative of $\chi^{2}$ (eq.~\ref{eq:chi2}) with respect to $k$. On the other hand, the parameter $l$ acts as a global calibration factor (zero-point) per band. This zero-point calibrates the offset between the observed fluxes and the best-fitting model and is defined as:
\begin{equation}\label{eq:li}
    l_{i} = \text{Median}[\phi_{i}^{\text{model}}/\phi_{i}^{\text{obs}}].
\end{equation}

In order to compute the photo-$z$, $\chi^{2}$ is minimized as a function of a set of template amplitudes on a redshift grid covering $0.01<z<2$, with a resolution of 0.001, so each galaxy has a corresponding best fitting template from which the redshift probability density $p(z)$ is computed. The photometric redshift that is assigned to each object corresponds to the peak of the $p(z)$ distribution, which we label as $z_{\rm{b}}$.

The templates we employ in our photo-$z$ estimation are the same as in \cite{Eriksen2019}. They include templates for elliptical and red spiral galaxies, star-bust galaxies are introduced following BC03 (\citealt{Bruzual}) models, with ages ranging from 0.03 Gyr to 3 Gyr. Additional BC03 templates with different ages and metallicites are also introduced. Emission lines are modeled for star-bust galaxies and their galactic extinction is accounted for following the Calzetti law (\citealt{Calzetti}).

\textsc{BCNZ} provides some photo-$z$ quality parameters that allow the galaxies with the best photo-$z$ to be selected, which might be advantageous for some applications. For a full description of the different quality parameters, see \cite{Eriksen2019}. Here, we have chosen to use the parameter $Q_{z}$, (eq.~\ref{eq:Qz}), since it is a combination of other quality parameters:

\begin{equation} \label{eq:Qz}
    Q_{z} \equiv \frac{\chi^{2}}{N_{f}-3}\left (\frac{z_{\textrm{quant}}^{99} - z_{\textrm{quant}}^{1}}{\textrm{ODDS}} \right ),
\end{equation}
where $\chi^{2}$ corresponds to eq.~\ref{eq:chi2}, $N_{f}$ is the number of filters, $z_{\textrm{quant}}^{n}$ are the nth percentile of the posterior distribution and ODDS is defined as:

\begin{equation}\label{eq:ODDS}
    \textrm{ODDS} = \int_{z_{\textrm{b}}-\Delta z}^{z_{\textrm{b}}+\Delta z} \textrm{d}z \; p(z),
\end{equation}
where $\Delta z = 0.035$. This last parameter quantifies the probability that is located around the peak of $p(z)$, that is, $z_{\textrm{b}}$.

We present a study of the performance of the PAUS wide fields as a function of $Q_{z}$ in Appendix~\ref{sec:Qz separation}, showing that the photo-$z$ quality is correlated with $Q_{z}$.

\subsection{New calibration of \texorpdfstring{$l$}{li}}
\label{sec:calibration}

One improvement that has been made to the \textsc{BCNZ} code is related to the estimation of $l$, the zero-point per band in eq.~\ref{eq:li}. In order to compute $l$, a comparison between the observed flux and the best fitting model is needed, so one has to evaluate the best fitting model over the whole redshift range. Since this is computationally expensive, in \cite{Eriksen2019} they defined a calibration sample for objects with spectroscopic redshifts and good photometry and only evaluated the best fitting model at the spectroscopic redshift of these objects, to later apply $l$ to the whole sample. We will refer to this method as the spectroscopic calibration. This restricts the technique to be applied only when spectroscopic redshift information is available. Another disadvantage of this method arises from the potential selection bias between the spectroscopic sample and the full catalogue. This could lead to an overestimation or underestimation of the proportion of objects, either in terms of magnitude or redshift, which might result in the application of a non-representative zero-point to the whole sample. Finally, the spectroscopic sample is used both to calibrate $l$ and to validate the photo-$z$, establishing a potential dependency in both steps. In order to alleviate these problems, we propose a new methodology to compute the zero-point.

Instead of computing the best fitting model at the spectroscopic redshift of each source in the calibration sample, the new technique uses an iterative approach. In the first iteration, $l$ is set to one for all the bands and the photo-$z$ are computed. In the next iteration, the best fitting model is evaluated at the photo-$z$ computed in the previous iteration and \textsc{BCNZ} is run again, setting l according to eq.~\ref{eq:li}. This iteration is performed five times in total, and it has been shown to be sufficient to reproduce the photo-$z$ accuracy obtained by the previous spectroscopic calibration, shown in Fig.~\ref{fig:iterative_vs_spectroscopic} (which will be discussed in Section~\ref{sec:Iterative Spectroscopic Method}). In order to better determine the zero-points, we also split the catalogue into 6 $i_{\textrm{AB}}$ bins containing equal numbers of objects and we compute the zero-points for each of these sub-samples. This way, we can better take into account the differences in the sample in terms of magnitude. We tried different approaches to divide the sample, such as using different numbers of magnitude and/or redshift bins, but found that the best solution is to divide it in 6 $i_{\textrm{AB}}$ bins.

The calibration sample used in this new iterative approach is defined so that the objects with poorer photometry are discarded. For each field, the survey mask of the reference catalogue is applied, stars are rejected and only objects observed in the 40 NB are kept for the calibration step. To reduce the computational time, we also downsample the calibration sample to consider 10\% of the objects in the whole catalogue. This is done except for the last iteration, where we apply $l$ to the full sample and compute all the photometric redshifts, extending the sample by not accounting for the survey mask and including objects previously classified as stars and with a coverage of at least 30 NB.

\begin{figure*}
    \centering
    \includegraphics[width=\textwidth]{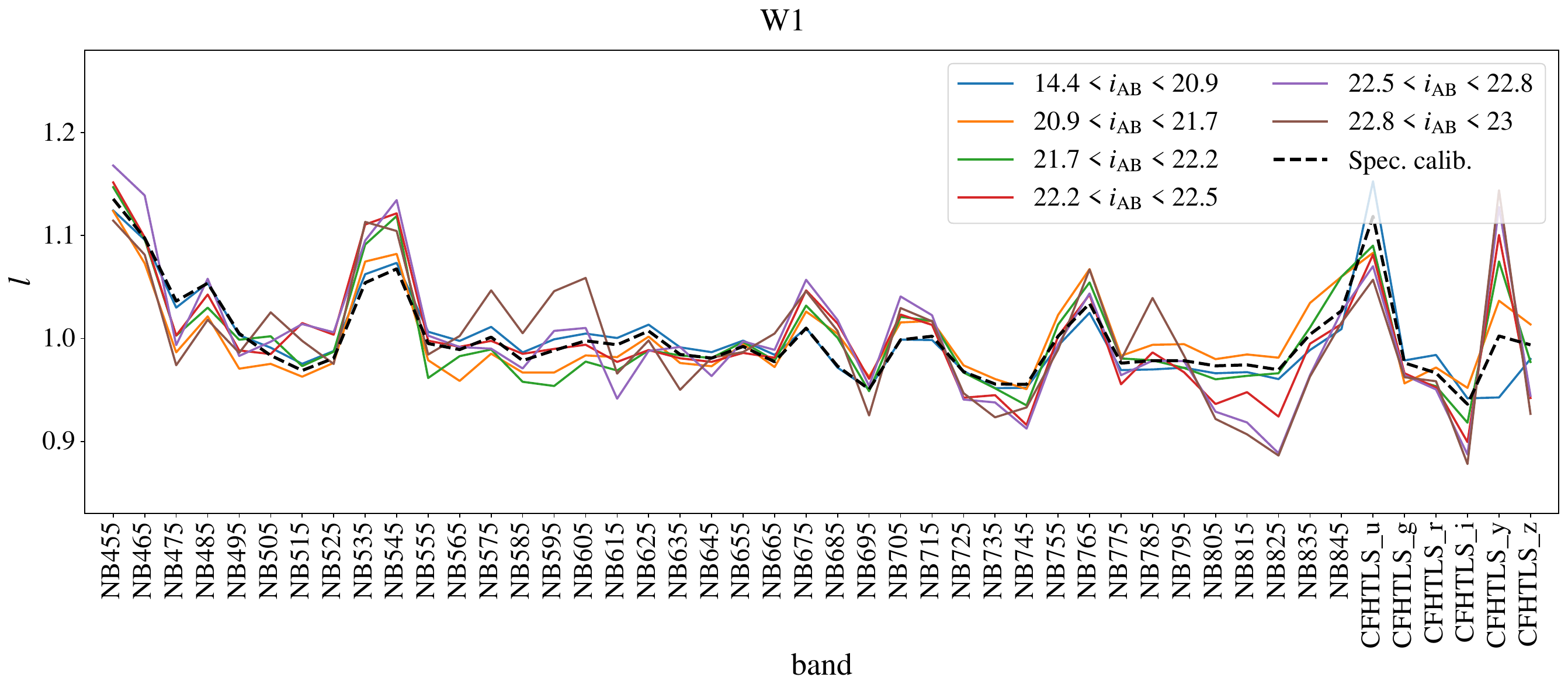}
    \caption{$l$ zero-points computed with the iterative (solid lines) and the spectroscopic (dashed line) calibration for the W1 field. The global trend for both cases is very similar. The spectroscopic calibration zero-point is much alike to the zero-point of the brightest bin in magnitude in the iterative calibration, since the majority of the objects in the spectroscopic calibration sample are in that range. In the iterative calibration method, variations in $l$ for the different $i_{\textrm{AB}}$ bins show that this method is able to capture the particular behaviour of each $i_{\textrm{AB}}$ subsample.}
    \label{fig:zp}
\end{figure*}

Fig.~\ref{fig:zp} shows the zero-points for the iterative (solid lines) and the spectroscopic (dashed line) calibration as a function of the 40 NB and the BB of the W1 field (similar trends are obtained for the other fields analysed). In the case of the iterative calibration, the distribution of the zero-points for the different magnitude bins follows a global trend, as expected, while preserving their differences. The peaks and off-peaks illustrate the corrections that need to be made in each band in order to fit the modelled flux. A similar global trend for the $l$ factor can be seen in the spectroscopic calibration method, which validates the new method. Note that the spectroscopic case closely follows the line corresponding to the brightest magnitude bin (blue line), since that is the bin containing the majority of objects in the spectroscopic calibration sample.

\subsection{Metrics}\label{section:Metrics}

The metrics that quantify the performance of the photometric redshifts are presented in this section. These metrics are based on $\Delta_{z}$, which is defined as:

\begin{equation}\label{eq:dx}
    \Delta_{z} = \frac{z_{\textrm{b}} - z_{\textrm{s}}}{1 + z_{\textrm{s}}},
\end{equation}
where $z_{\textrm{b}}$ and $z_{\textrm{s}}$ correspond to the photometric and spectroscopic redshifts, respectively. This quantity quantifies how accurately the photo-$z$ are estimated in comparison with the spec-$z$. Note that the metrics associated with $\Delta_{z}$ will only be available for objects with spec-$z$ information.

We define $\sigma_{68}$ of the quantity $\Delta_{z}$ as:
\begin{equation}
    \sigma_{68} = \frac{P[84]-P[16]}{2},
\end{equation}
where P[$x$] corresponds to the percentile $x$ of the $\Delta_{z}$ distribution.

In this work, an object is considered to be an outlier if it satisfies the following condition:
\begin{equation}\label{eq:outlier}
    \left | \Delta_{z} \right |> 0.1.
\end{equation}

Finally, the bias shows the systematic difference between the spectroscopic and  photometric redshifts and is defined as the median of the difference:

\begin{equation}\label{eq:bias}
    \mu = \mathrm{med}(z_{\textrm{b}}-z_{\textrm{s}}).
\end{equation}

The centralised scatter, $\sigma_{68}$, and the outlier fraction of the quantity $\Delta_{z}$, and the bias of $z_{\textrm{b}} - z_{\textrm{s}}$ will be quantified as a function of the $i_{\textrm{AB}}$, the spectroscopic and the photometric redshift.

\subsection{Weighted photo-\texorpdfstring{$z$}{z}}
\label{sec:Weighted photo-$z$}

\begin{figure*}
    \centering
    \includegraphics[width=0.85\textwidth]{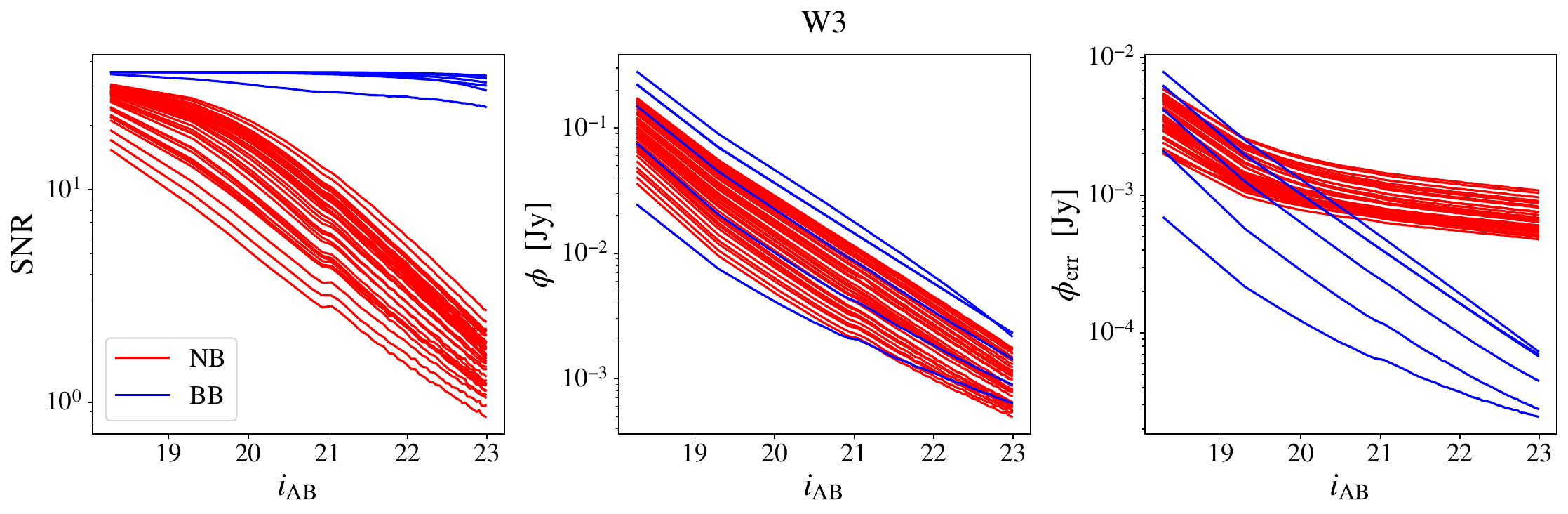}
    \caption{SNR, flux ($\phi$) and flux error ($\phi_{\mathrm{err}}$) (from left to right) of the W3 field for the broad bands (blue lines) and the narrow bands (red lines). The SNR of the broad bands stays fairly constant for the entire $i_{\textrm{AB}}$ range, since both the BB flux and flux errors continuously decrease. In the case of the narrow bands, the SNR decreases as the objects become fainter, which is caused by the constant NB flux errors at faint magnitudes.}
    \label{fig:SNR_weight_W3}
\end{figure*}

We obtain that, for very faint objects ($i_{\textrm{AB}}> 22.5$), the performance of the photo-$z$ computed using the PAUS NB+BB does not improve compared to the photo-$z$ computed using BB photometry only. This is explained by the low SNR of the PAUS photometry for these faint objects. This low SNR is due to the fact that the noise in the NB is dominated by (Poisson) sky noise above $i_{\textrm{AB}}> 22.5$, whereas the BB fluxes correspond to much deeper exposures. Additionally, the BB collect light over a wavelength range an order of magnitude higher, integrating more signal. All this is illustrated by Fig.~\ref{fig:SNR_weight_W3}, where the SNR, the flux ($\phi$) and the flux error ($\phi_{\mathrm{err}}$) for the W3 field are shown (the case of W1 is similar, although not shown here for brevity). On the one hand, the errors in the NB flux stay almost constant for objects fainter than $i_{\textrm{AB}}\sim20$, while the errors in the BB decrease over the full $i_{\textrm{AB}}$ range. On the other hand, the fluxes for both NB and BB continuously decrease as a function of $i_{\textrm{AB}}$. This combination makes the SNR of the NB to decrease as we reach fainter magnitudes, dropping to SNR values $\sim$1-3 beyond $i_{\textrm{AB}}\sim22.5$, while the SNR of the BB stays almost constant. For the brighter objects, the slope of the narrow bands SNR is shallower because there are more objects measured as extended. The case for the G09 field is slightly different regarding the BB flux errors. We include a study on the different SNR and photo-$z$ performance between the W1/W3 and the G09 fields in Appendix~\ref{sec:Study fields}.

\begin{figure*}
    \centering
    \includegraphics[width=0.8\textwidth]{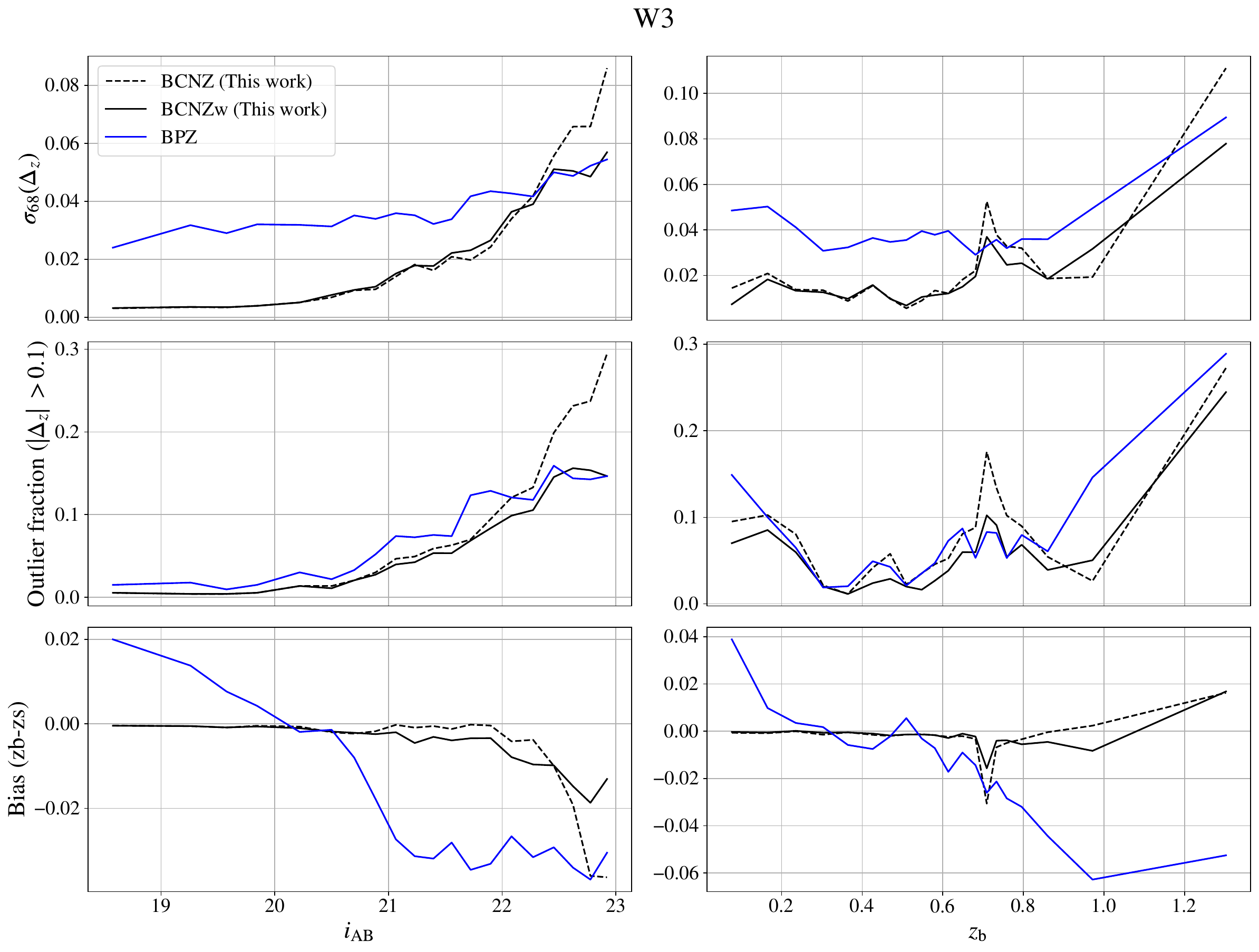}
    \caption{$\sigma_{68}$ (top), outlier fraction (middle) and bias (bottom) as a function of $i_{\textrm{AB}}$ (left) and $z_{\textrm{b}}$ (right) for the photometric redshifts in the W3 field. The $z_{\textrm{b}}$ computed from \textsc{BPZ} (using only BB) are shown in blue, the ones computed by \textsc{BCNZ} (using NB+BB) in dashed black lines and the weighted combination of \textsc{BCNZ} and \textsc{BPZ} photo-$z$ (\textsc{BCNZ}w) in solid black lines. Each bin is defined to contain an equal number of objects. For bright objects, $z_{\textrm{b, \textsc{BCNZ}}}$ have better performance in terms of smaller values of $\sigma_{68}$, outlier fraction and bias. However, for fainter objects ($i_{\textrm{AB}}>22.5$), $z_{\textrm{b, \textsc{BPZ}}}$ has better accuracy in terms of $\sigma_{68}$ and outlier fraction. The weighted combination of \textsc{BCNZ} and \textsc{BPZ} yields the best overall photo-$z$ performance in terms of $\sigma_{68}$, outlier fraction and bias.}
    \label{fig:weighted_zb}
\end{figure*}

In order to see how the differences in SNR between NB and BB affect the estimation of the photo-$z$, the top panels of Fig.~\ref{fig:weighted_zb} show the $\sigma_{68}$ as a function of $i_{\textrm{AB}}$ and $z_{\textrm{b}}$ for the photo-$z$ estimated from NB+BB (dashed black line) and BB-only (blue line) for the W3 field (similar trends are observed for the W1 and G09 fields). At high SNR (i.e. for bright galaxies) the accuracy depends linearly on the width of the filters, reducing the $\sigma_{68}$ by a factor of 10 (from few $10^{-2}$ to few $10^{-3}$) when using NB, which are a factor of 10 narrower than the BB. Also, note how the left panel of Fig.~\ref{fig:SNR_weight_W3} and the top left panel of Fig.~\ref{fig:weighted_zb}, which respectively show the SNR in the different filters and the $\sigma_{68}$ as a function of $i_{\textrm{AB}}$, look like the reverse of each other. This indicates that the degradation of the photo-$z$ performance with magnitude is mostly driven by the SNR decrease. 
We have verified that this is indeed the case by doing different data reductions with a different number of exposures (see Fig. 26 in \citealt{Serrano2022}).

It is important to highlight that the BB photo-$z$ were estimated by the CFHTLenS and KiDS teams (\citealt{CFHTLs_photoz}, \citealt{KiDS_DR4}) using a different photo-$z$ code (\textsc{BPZ}, \citealt{Benitez2000} ) which was optimized for BB filters (for a detailed comparison with \textsc{BCNZ} see \citealt{Eriksen2019}). We will refer to the photo-$z$ computed using only BB as $z_{\textrm{b, \textsc{BPZ}}}$, and the photo-$z$ computed with NB+BB as $z_{\textrm{b, \textsc{BCNZ}}}$. Even when the BB and NB+BB estimates share some of the same data, the choice of templates, method and optimization makes the two estimates fairly independent. This is illustrated in  Fig.~\ref{fig:weighted_zb}, which also shows the inverse variance weighted photo-$z$ (solid black line), combining the BB and NB+BB cases, defined as:
\begin{equation}\label{eq:zb_weighted}
    z_{\textrm{b, \textsc{BCNZ}w}} = \frac{z_{\textrm{b, \textsc{BCNZ}}}\cdot w_{\textrm{\textsc{BCNZ}}} + z_{\textrm{b, \textsc{BPZ}}}\cdot w_{\textrm{\textsc{BPZ}}}}{w_{\textrm{\textsc{BCNZ}}}+w_{\textrm{\textsc{BPZ}}}},
\end{equation}
where the weight $w = 1/\sigma_{68, i_{\textrm{AB}}}^{2}$ is given by the corresponding $\sigma_{68}$ values of the photo-$z$ of \textsc{BCNZ} or \textsc{BPZ} as a function of $i_{\textrm{AB}}$ in Fig.~\ref{fig:weighted_zb}.
The weighting of $z_{\textrm{b, \textsc{BCNZ}}}$ and $z_{\textrm{b, \textsc{BPZ}}}$ allows us to obtain a new $z_{\textrm{b, \textsc{BCNZ}w}}$, which closely follows the scatter of the photo-$z$ from $z_{\textrm{b, \textsc{BCNZ}}}$ for objects brighter than $i_{\textrm{AB}}= 22.5$. When the performance of $z_{\textrm{b, \textsc{BCNZ}}}$ substantially decreases, $z_{\textrm{b, \textsc{BCNZ}w}}$ follows the performance of $z_{\textrm{b, \textsc{BPZ}}}$. Also, at the intersection between the NB+BB and the BB cases, the new $z_{\textrm{b, \textsc{BCNZ}w}}$ shows slightly better $\sigma_{68}$ than either of the other two cases.

The bottom panel in Fig.~\ref{fig:weighted_zb} shows the bias as a function of $i_{\textrm{AB}}$ and as a function of $z_{\textrm{b}}$. We note that $z_{\textrm{b, \textsc{BPZ}}}$ presents a larger bias than $z_{\textrm{b, \textsc{BCNZ}}}$. In order not to transfer that bias into our weighted estimate, we subtract the bias from $z_{\textrm{b, \textsc{BPZ}}}$ before applying the inverse weighting technique. By doing that, we end up with a similar bias between $z_{\textrm{b, \textsc{BCNZ}}}$ and $z_{\textrm{b, \textsc{BCNZ}w}}$.

\begin{figure*}
    \centering
    \includegraphics[width=1.\textwidth]{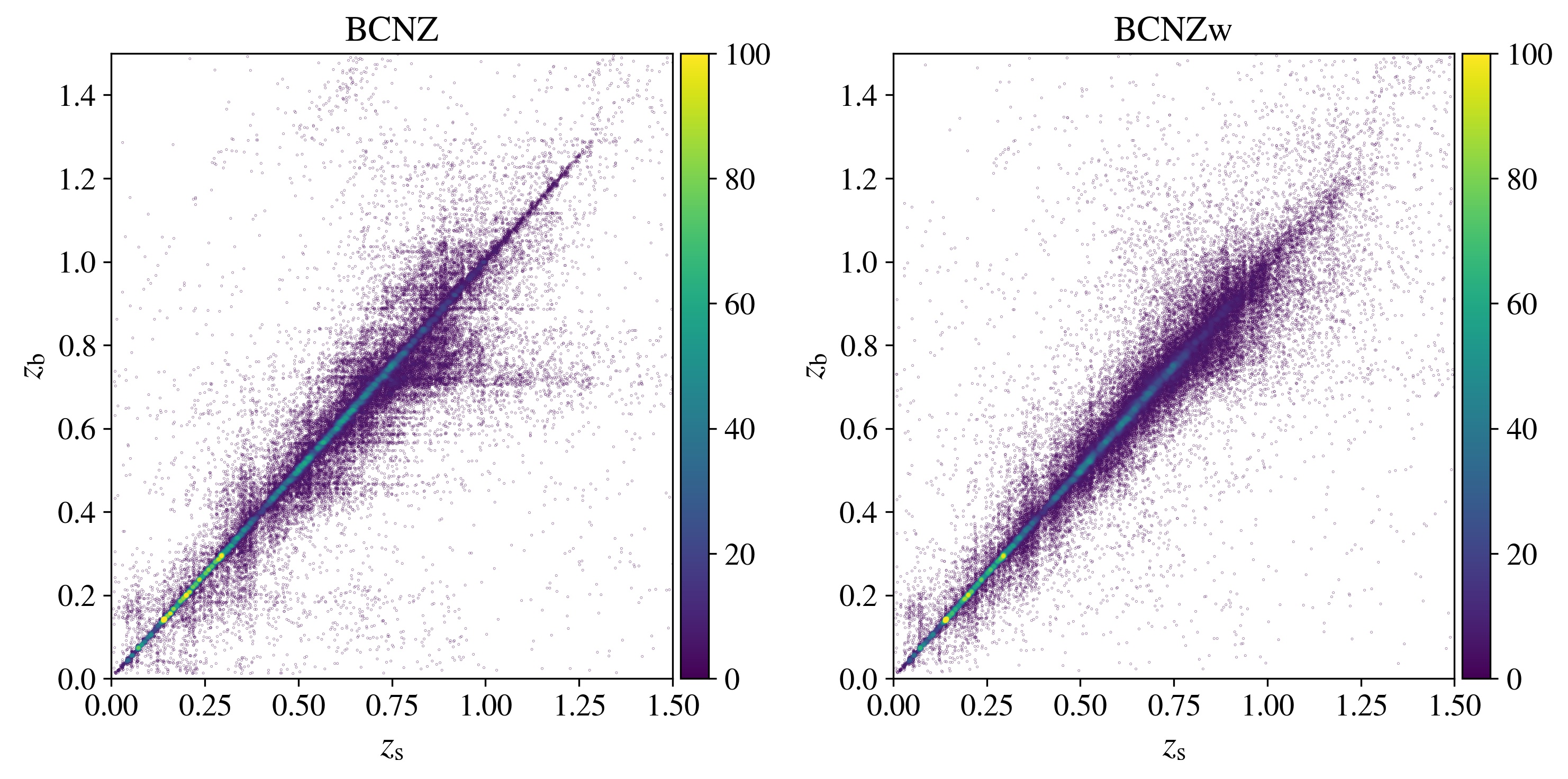}
    \caption{Photometric redshift vs spectroscopic redshift for the \textsc{BCNZ} photo-$z$ (left) and the \textsc{BCNZ}w photo-$z$ (right), that is, when the inverse variance weighting is applied, for the 3 PAUS wide fields combined. The colour bar indicates the density of objects. The horizontal stripes at $z_{\textrm{b}} \approx 0.72$ and $z_{\textrm{b}} \approx 0.89$ are dissipated when weighting with the \textsc{BPZ} photo-$z$, computed only with broad bands.}
    \label{fig:weighted_zb_vs_zs}
\end{figure*}

Fig.~\ref{fig:weighted_zb_vs_zs} shows the scatter plot of photo-$z$ vs spec-$z$ of the three PAUS wide fields for the two photo-$z$ versions, unweighted (\textsc{BCNZ}, left) and weighted (\textsc{BCNZ}w, right), of the NB+BB estimates. An issue that is solved by the inverse weighted photo-$z$ is the fact that, a small fraction ($\sim$$2\%$ in the W1 and W3 fields and $\sim$$3\%$ in the G09 field) of the photo-$z$ computed with \textsc{BCNZ} with $z_{\textrm{s}} > 0.75$ for W1 and W3 and $z_{\textrm{s}} > 0.85$ for G09, have a wrongly assigned value close to $z_{\textrm{b}}$ $\simeq  0.72$ or $z_{\textrm{b}}$ $\simeq 0.89$, respectively. This creates a horizontal stripe (or focusing) in the $z_{\textrm{b}}$ vs. $z_{\textrm{s}}$  plots (left panel in Fig. \ref{fig:weighted_zb_vs_zs} and upper panels in Fig. \ref{fig:weighted_zb_vs_zs_fields} to see each field in detail). 
This redshift focusing effect occurs when the prior dominates the posterior probability distribution in the low SNR case. In this case, a large number of objects are assigned the peak value of the prior, which leads to artificial peaks in the redshift histograms (\citealt{CFHTLs_photoz}). Even though some small focusing effect is expected when computing photo-$z$, the cases at $z_{\textrm{b}}$ $\simeq  0.72$ and $z_{\textrm{b}}$ $\simeq  0.89$ become an issue for us. Because of the different priors and libraries of SED and emission lines (see \citealt{Eriksen2019} for details), the focusing effect is different in the two photo-$z$ codes. Therefore, the focusing issue is much dissipated in the newly defined $z_{\textrm{b, \textsc{BCNZ}w}}$ (compare left and right panels of Fig.~\ref{fig:weighted_zb_vs_zs}). 

However, for the G09 and the KiDZ-COSMOS fields, the peak of objects at $z_{\textrm{b}}$ $\simeq 0.89$ is not completely dissipated by $z_{\textrm{b, \textsc{BCNZ}w}}$, so that an artificial excess of objects at that redshift remains. In the case of the G09 field, we found that some of those outliers correspond to objects for which the BB photo-$z$ are higher than 4, so we removed from the sample all objects with $z_{\textrm{b, \textsc{BPZ}}} > 4$, which accounted for $\sim$$1 \%$ of the G09 objects. The remaining outliers ($\sim$$2 \%$ and $\sim$$1 \%$ of the total sample in the G09 and KiDZ-COSMOS fields, respectively) were not removed, but instead their photometric redshift were substituted by the BB photo-$z$ provided by KiDS. Since those redshifts have a resolution of 0.01 and \textsc{BCNZ} has a resolution of 0.001, we applied a normal random distribution with a $\sigma = 0.01$ to disperse them around their initial value, matching the resolution of \textsc{BCNZ}-like photo-$z$.

Taking these two factors into account, that is, the low SNR in the NB for faint objects and the horizontal stripe around $z_{\textrm{b, \textsc{BCNZ}}} \sim 0.72$ and $z_{\textrm{b, \textsc{BCNZ}}} \sim 0.89$, we will study the performance of the \textsc{BCNZ} photo-$z$ and the \textsc{BCNZ} weighted photo-$z$ and compare them.

\subsection{Validating the G09 field with KiDZ-COSMOS}\label{sec:KIDZ_COSMOS}

As mentioned in Section~\ref{sec:Spectroscopic}, the validation sample in the G09 field is not adequate neither in terms of $i_{\textrm{AB}}$ nor spec-$z$ coverage, and is therefore very limited to assess the performance of the G09 photo-$z$. For that reason, a new validation sample is introduced named KiDZ-COSMOS, which covers a range of $i_{\textrm{AB}}$ magnitude and spec-$z$ not covered before. These objects lie in the COSMOS field and were observed by KiDS, so their photometry is equivalent to the objects observed in the G09 field.

For this test, the positions of the objects in COSMOS are defined by using forced photometry with the KiDZ-COSMOS catalogue, which allows us to measure the PAUS NB fluxes and their errors. Although in the COSMOS field PAUS usually does an average of 5 single exposures before coadding, for the purpose of this test we limit the average number of single exposures to 3, since this is the average number for the PAUS wide fields. Once the forced photometry is done, the photo-$z$ are measured using the NB and BB (\textit{ugrizYJHK$_{s}$}) from KiDZ-COSMOS.

\begin{figure}
    \centering
    \includegraphics[width=0.4\textwidth]{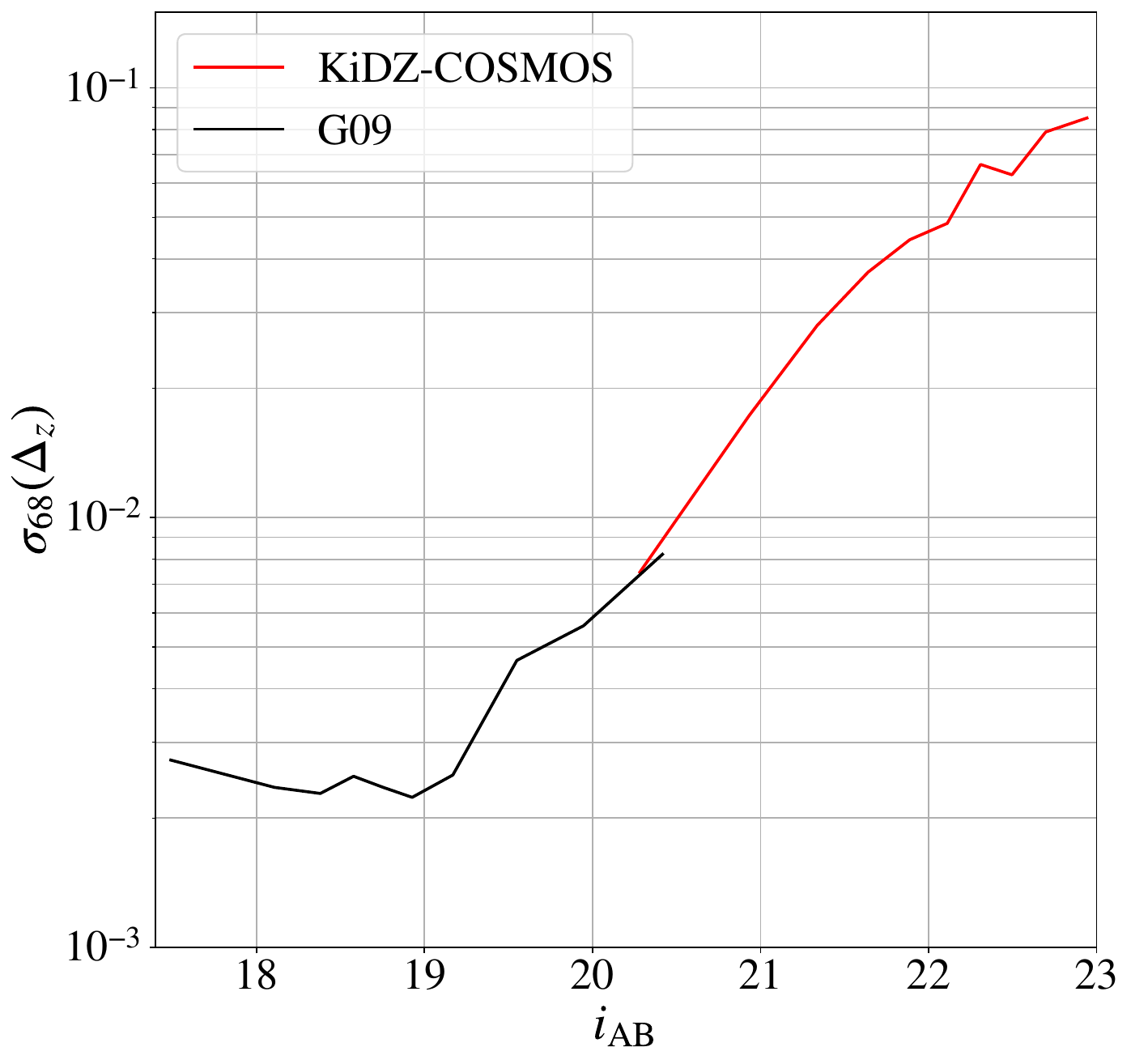}
    \caption{$\sigma_{68}(\Delta_{z})$ as a function of $i_{\textrm{AB}}$ for the G09 and the KiDZ-COSMOS validation samples. While G09 can be used to validate the bright end of $i_{\textrm{AB}}$, KiDZ-COSMOS validates the faintest objects. In the magnitude range where both validation samples meet ($i_{\textrm{AB}} \sim 20.5$), the $\sigma_{68}$ of both cases intersect, showing similar performances.}
    \label{fig:kidz_cosmos_performance}
\end{figure}

Fig.~\ref{fig:kidz_cosmos_performance} shows the performance of the photometric redshifts as a function of $i_{\textrm{AB}}$ for the G09 and the KiDZ-COSMOS validation samples. Since the $i_{\textrm{AB}}$ range of both samples is complementary (G09 covers most of the bright objects, while KiDZ-COSMOS covers the faintest), each one allows us to determine the performance for different magnitude ranges. At the magnitude where both validation samples meet, they present similar $\sigma_{68}$ values, as expected for objects observed with the same photometric survey.

Finally, from now on we will use the G09 + KiDZ-COSMOS validation sample as the standard for G09 and will refer to it only as the G09 validation sample, unless indicated otherwise.

\section{Photo-\texorpdfstring{$z$}{z} catalogues}
\label{sec:Photo-z catalogues}

In this section, the photometric catalogues are presented in detail. First, in Section~\ref{sec:Iterative Spectroscopic Method}, the new calibration technique is validated. Later, in Section~\ref{sec:Comparison Validation}, a comparison of the radial distribution of the PAUS wide fields and the performance of the photo-$z$ are shown. Next, in Section~\ref{sec:colour separation} a detailed analysis is performed as a function of the galaxy colours. Finally, in Section~\ref{sec:Performance of the $p(z)$} we study the $p(z)$ distribution obtained by \textsc{BCNZ}.

We publish the photo-$z$ studied in this work in CosmoHub (\citealt{CosmoHub1, CosmoHub2}), where they can be accessed under reasonable demand to the authors. Table \ref{tab:photo-z_publish} shows the name and the description of the columns included in this catalogue.

\subsection{Iterative vs. spectroscopic method}
\label{sec:Iterative Spectroscopic Method}

In Section~\ref{sec:calibration}, we introduced a new calibration technique of the zero-point $l$ (see eq.~\ref{eq:li}) that consisted of an iterative approach, where $l$ was set to 1 in the first iteration and updated by the photometric redshifts computed in the previous iteration. This method was in contrast with the earlier one used by \textsc{BCNZ}, where the calibration was performed via a subset with spectroscopic redshifts. Thus, this new technique allows to calibrate the photo-$z$ estimation even when there are not spectroscopic redshifts available. It is also important to note that the previous calibration method used the same objects with spectroscopic redshifts to calibrate and validate the photo-$z$, so that both steps of the photo-$z$ estimation process were not independent.

\begin{figure}
    \centering
    \includegraphics[width=0.4\textwidth]{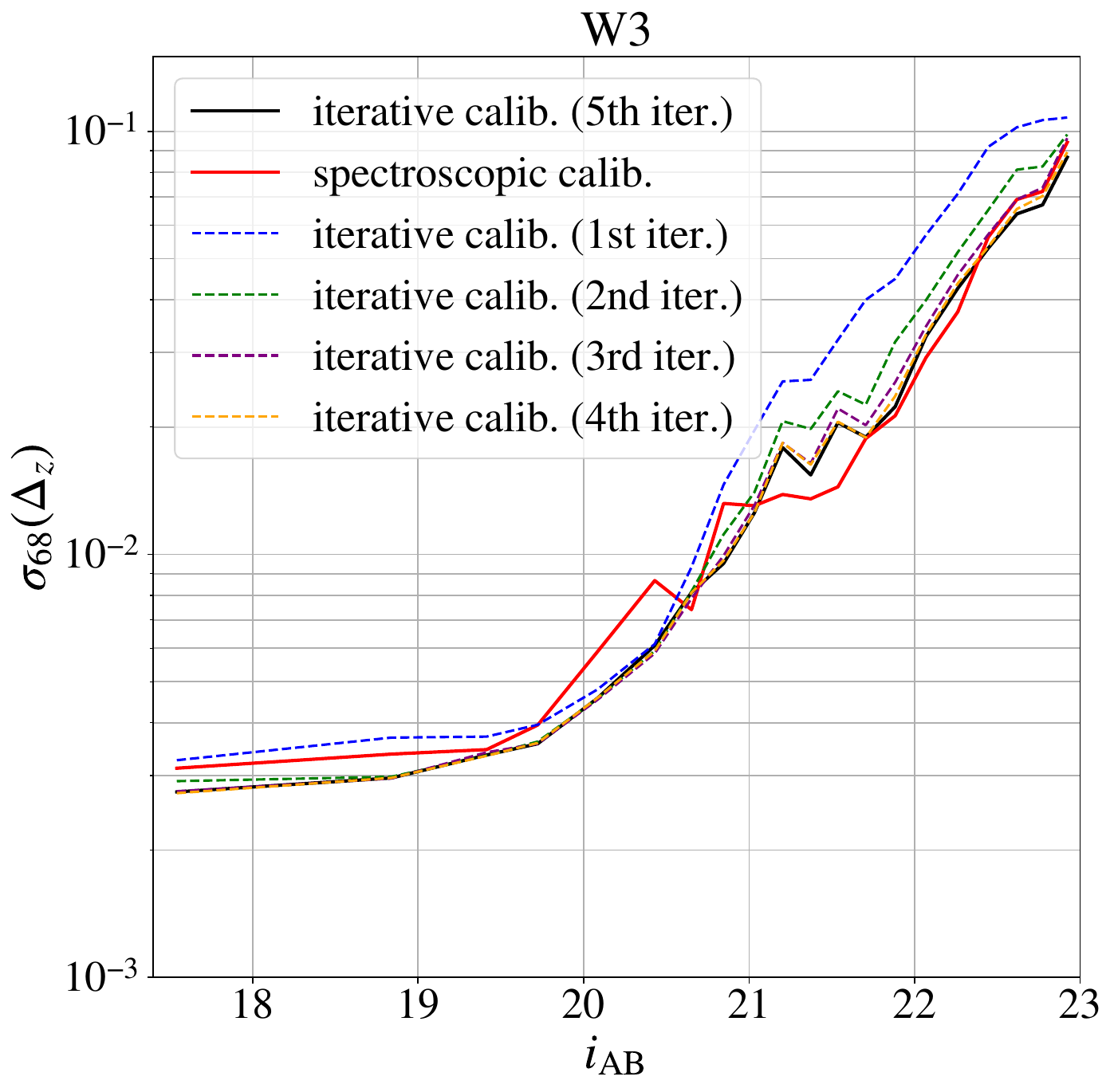}
    \caption{Comparison of the $\sigma_{68}$ as a function of $i_{\textrm{AB}}$ for the iterative and the spectroscopic calibration. Solid lines show the final $\sigma_{68}$ for the iterative and the spectroscopic calibration, while dashed lines show how the intermediate results for the iterative calibration converge to the final iteration. The performance achieved by both techniques is equivalent.}
    \label{fig:iterative_vs_spectroscopic}
\end{figure}

To ensure that this iterative technique gives, at least, the same performance as the spectroscopic one, we show the $\sigma_{68}(\Delta_{z})$ for both methods in the W3 field in Fig.~\ref{fig:iterative_vs_spectroscopic}, where the solid lines are the final results of both calibration techniques and the dashed lines are the intermediate results computed at each iteration in the iterative technique. From the solid lines, it is noticeable that both methods agree in performance, since the $\sigma_{68}$ values are equivalent for the whole $i_{\textrm{AB}}$ range. On the other hand, the dashed lines indicate that, iteration by iteration, the iterative calibration converges to the final result, while also indicating that 5 iterations are sufficient for convergence.

\subsection{PAUS wide fields photometric redshifts comparison}
\label{sec:Comparison Validation}

The aim of this section is to present the radial distribution of the PAUS wide fields photo-$z$ catalogues, study the photo-$z$ performance and compare the different estimates of the photo-$z$, that is, $z_{\textrm{b, \textsc{BCNZ}}}$ and $z_{\textrm{b, \textsc{BCNZ}w}}$. 

The analysis is done after excluding the objects with bad photometry or those classified as stars, following the flags and masks defined on Section~\ref{sec:BBphot}. Also, \textsc{BCNZ} allows to compute photometric redshifts for objects with different NB coverages and we opt to use objects with a coverage of 30 NB or more, as their performance is very similar to that obtained using objects with a coverage of only 40 NB (see Appendix~\ref{sec:NB coverage} for more details). As a result, we gain almost 300\,000 objects with a coverage below 40 NB. Finally, as detailed in Section~\ref{sec:Weighted photo-$z$}, some objects of the G09 field with $z_{\textrm{b, \textsc{BPZ}}}>4$ are considered outliers, so we remove them from the catalogue.

Table~\ref{tab:Photo-z} shows the number of objects with photo-$z$ information, the masked area, the number density and the average $\sigma_{68}$ for both photo-$z$ estimates, \textsc{BCNZ} and \textsc{BCNZ}w. The number densities attained by PAUS are much higher than those reached with the spectroscopic surveys defined in Section~\ref{sec:Spectroscopic}, allowing us to have high quality redshift information between broad band photometric and spectroscopic redshifts for high density regions on the sky. The mean $\sigma_{68}$ of both estimates is essentially the same, although the differences appear when studying it as a function of $i_{\textrm{AB}}$, $z_{\textrm{b}}$ and $z_{\textrm{s}}$, as it will be discussed in detail in Fig.~\ref{fig:metrics_all_WFs}.

\begin{table}
\caption{Number of photometric redshifts, masked area, number density and average $\sigma_{68}$ of the $z_{\textrm{b}}$ \textsc{BCNZ} and \textsc{BCNZ}w for the PAUS wide fields observed by PAUS. These values correspond to objects observed with at least 30 NB, after applying the mask and rejecting stars.}
\begin{center}
\label{tab:Photo-z}
\begin{tabular}{c c}
\hline
\hline
\# masked photo-$z$ & 1241047 \\ \hline
Masked area [deg$^{2}$] & 40.99 \\ \hline
Number density [deg$^{-2}$] & 30277 \\ \hline
\textsc{BCNZ} $\sigma_{68}$ & 0.019 \\ \hline
\textsc{BCNZ}w $\sigma_{68}$ & 0.020 \\ \hline
\end{tabular}
\end{center}
\end{table}

\begin{figure}
    \centering
    \includegraphics[width=0.4\textwidth]{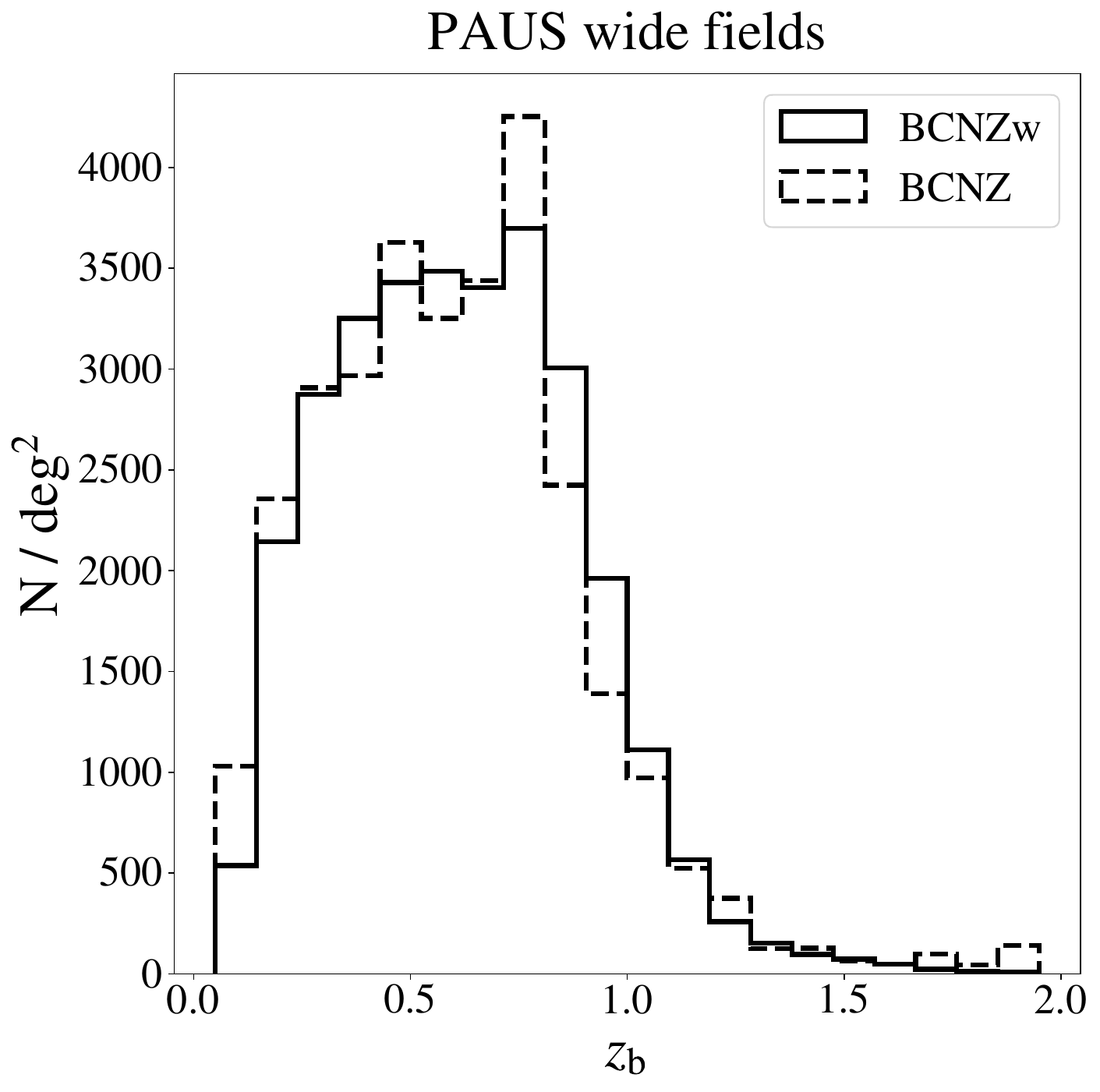}
    \caption{Photometric redshift distributions divided by the area as a function of $z_{\textrm{b}}$ for the PAUS wide fields. The output of \textsc{BCNZ} and the inverse variance weighting (\textsc{BCNZ}w) cases are shown. The \textsc{BCNZ}w photo-$z$ do not show the artificial peaks seen in the redshifts from \textsc{BCNZ}.}
    \label{fig:nz_wide_fields_combined}
\end{figure}

The photometric redshift distributions divided by the area as a function of the photometric redshifts are presented in Fig.~\ref{fig:nz_wide_fields_combined} for the combined PAUS wide fields, for the \textsc{BCNZ} original photo-$z$ (dashed black line) and for the weighted cases (solid black line). It can be seen that there are some artificial peaks around $z_{\textrm{b}}\sim0.7-0.8$ in the $z_{\textrm{b, \textsc{BCNZ}}}$, wrongly assigned by the photo-$z$ code, as discussed in Section~\ref{sec:Weighted photo-$z$}. By weighting with the BB photo-$z$, these artificial peaks are mostly removed, leading to a smoother distribution.

\begin{figure*}
    \centering
    \includegraphics[width=0.95\textwidth]{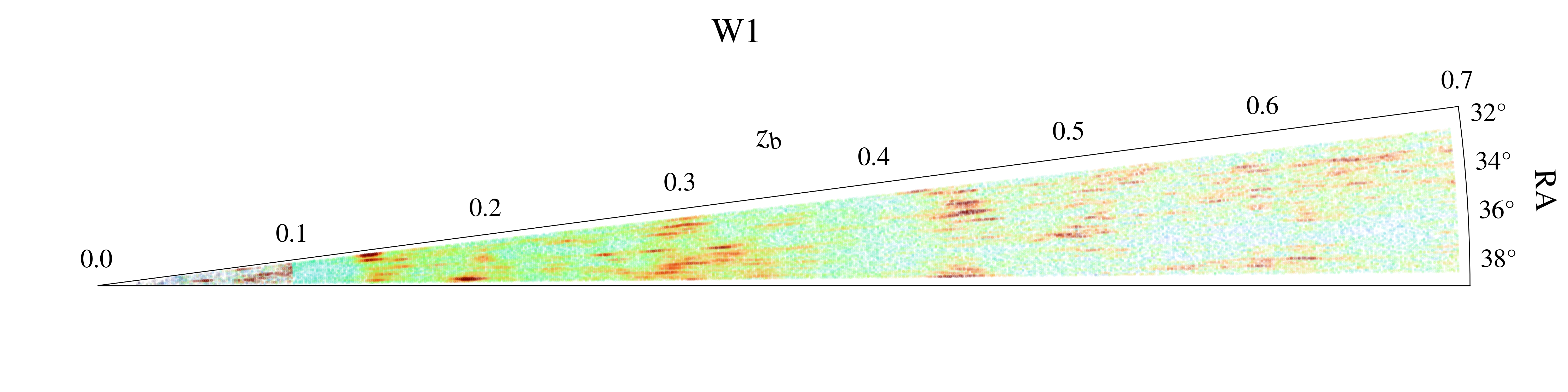}
    \includegraphics[width=0.95\textwidth]{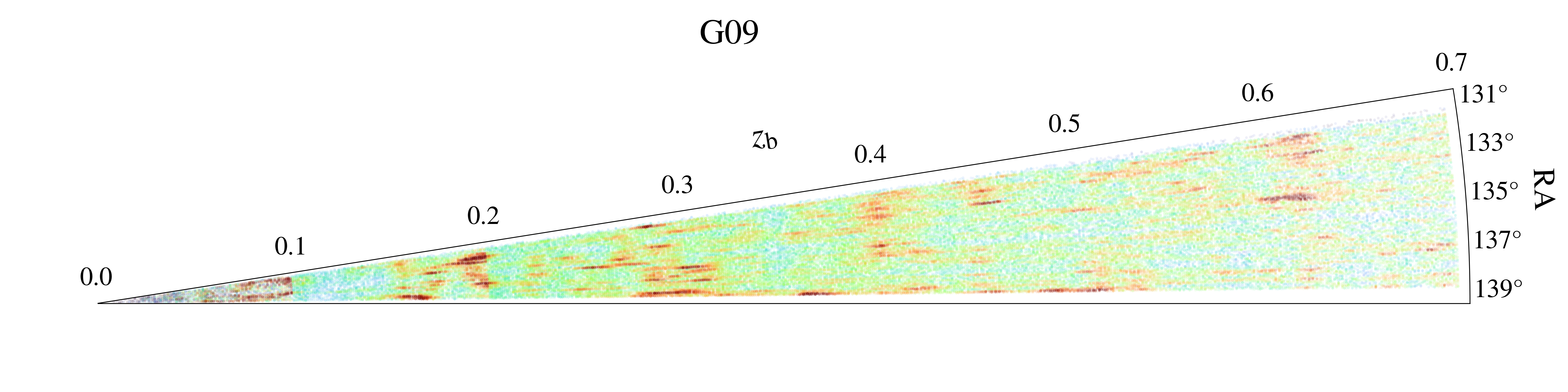}
    \includegraphics[width=0.95\textwidth]{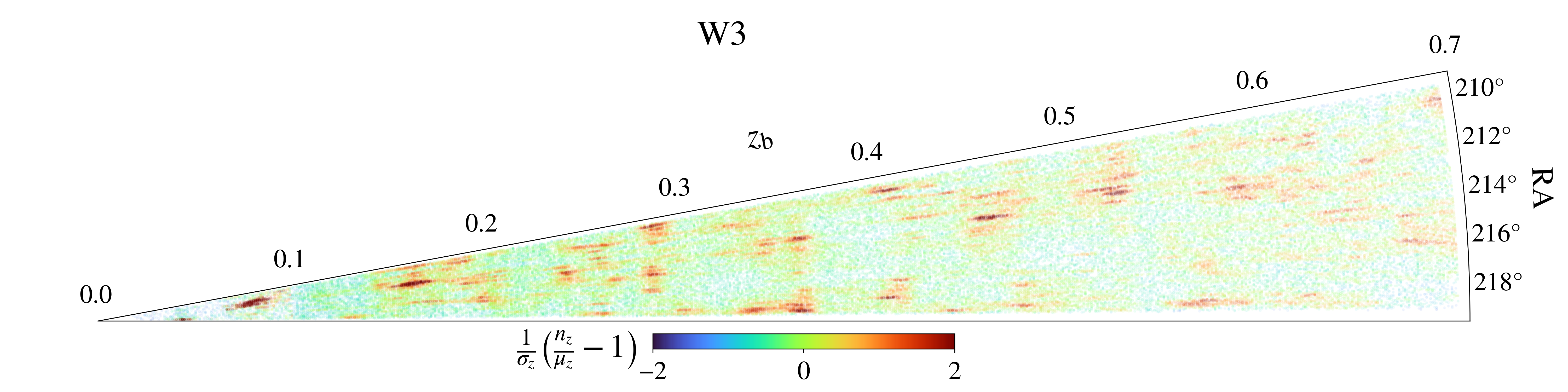}
    \caption{Distribution of the photometric redshifts of the W1, G09 and W3 fields as a function of RA for a cut in DEC, selecting the best 50\% of objects based on $Q_{z}$. The colour bar indicates the fluctuation in the number density of objects per redshift bin ($n_{z}$), $\frac{1}{\sigma_{z}}\left ( \frac{n_{z}}{\mu_{z}}-1\right )$, such that $\sigma_{z}$ and $\mu_{z}$ correspond to the $\sigma_{68}$ and the median of $n_{z}$.}
    \label{fig:LSS_WF}
\end{figure*}

\begin{figure*}
    \centering
    \includegraphics[width=0.95\textwidth]{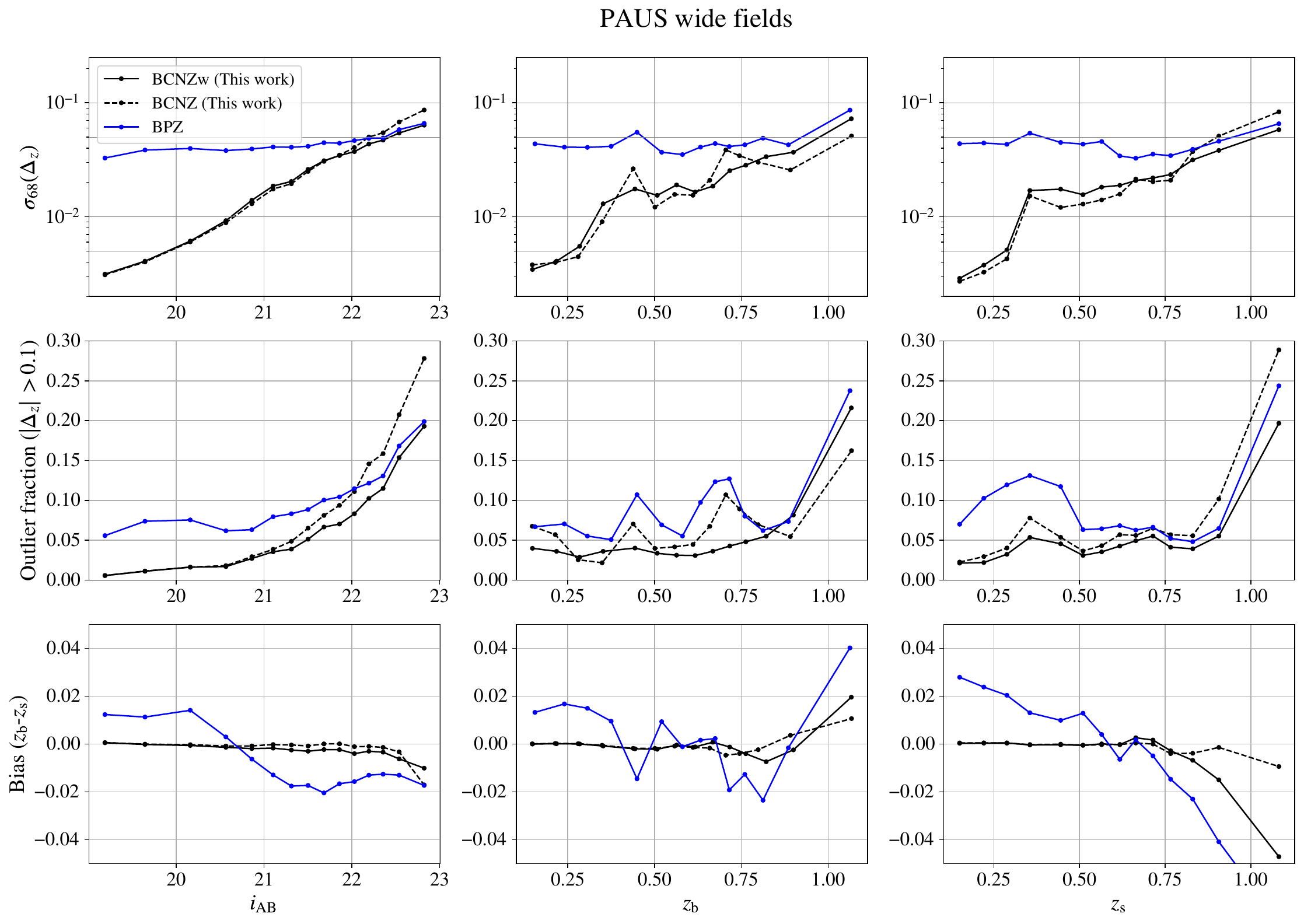}
    \caption{Performance of the PAUS wide fields for the weighted \textsc{BCNZ}w photo-$z$ (solid black line), the \textsc{BCNZ} photo-$z$ (dashed black line) and the broad band photo-$z$ computed from \textsc{BPZ} (blue line). From top to bottom, the $\sigma_{68}$, the outlier fraction and the bias are shown as a function of $i_{\textrm{AB}}$ magnitude, photometric redshift and spectroscopic redshift. Each bin is defined to contain an equal number of objects. In general, BCNZ photo-$z$ estimates give superior performance in all metrics considered, in comparison with BPZ photo-$z$.}
    \label{fig:metrics_all_WFs}
\end{figure*}

Fig.~\ref{fig:LSS_WF} shows the distribution of the weighted photometric redshifts for a subsample of 2 deg$^{2}$ in DEC as a function of RA, coloured by the fluctuation of the number density of objects for all three wide fields. This fluctuation in density is defined as $\frac{1}{\sigma_{z}}\left ( \frac{n_{z}}{\mu_{z}}-1\right )$, where $\sigma$ and $\mu$ correspond, respectively, to the $\sigma_{68}$ and the median value of the number density, $n$, and the subscript $z$ reflects that we slice in redshift bins. In this case, we compute the quantities in redshift bins, as opposed to the case in Fig.~\ref{fig:Area_density}, since we want to focus in the fluctuations at each redshift. In order to select the best objects, we keep the 50\% of objects with better photo-$z$ quality based on the $Q_{z}$ parameter. These figures illustrate the good determination of the photo-$z$ in PAUS, highligting lower and higher overdensity regions. We can see some photo-$z$ errors around some of the overdense regions, which can be identified at fixed RA as a function of redshift. This can be seen, for example, in the W1 field at RA $\sim$38° and $z_{\textrm{b}}$$\sim0.2$ or in the G09 field at RA $\sim$138° and $z_{\textrm{b}}$$\sim0.3$.

Fig.~\ref{fig:metrics_all_WFs} shows the photo-$z$ performance in the PAUS wide fields as a function of $i_{\textrm{AB}}$, $z_{\textrm{b}}$ and $z_{\textrm{s}}$. From top to bottom, the $\sigma_{68}$, the outlier fraction and the bias are shown. Solid black lines show the \textsc{BCNZ} weighted photo-$z$ (\textsc{BCNZ}w), dashed black lines show the \textsc{BCNZ} photo-$z$ and solid blue lines show the \textsc{BPZ} photo-$z$. As shown before in Fig.~\ref{fig:weighted_zb}, the $\sigma_{68}$ as a function of $i_{\textrm{AB}}$ is lower for faint objects for the weighted photo-$z$ case than for the \textsc{BCNZ} photo-$z$; the improvement in $\sigma_{68}$ of the  BCNZw photo-$z$ starts to become apparent at $i_{\textrm{AB}} > 22.5$, near where the \textsc{BCNZ} and the \textsc{BPZ} lines cross. However, this is not necessarily the case as a function of $z_{\textrm{b}}$ or $z_{\textrm{s}}$, where the $\sigma_{68}$ does not show a clear preference for either of the \textsc{BCNZ} estimates. For the weighted photo-$z$, $\sigma_{68}$ ranges between $\sim$0.003 and $\sim$0.06 for $i_{\textrm{AB}} \sim 19$ and 23, respectively. Meanwhile, the \textsc{BCNZ} case drops in value to $\sigma_{68}\sim0.09$ at the faintest magnitude bin. As for the outlier fraction, we note that, for all the $i_{\textrm{AB}}$ and most of the $z_{\textrm{b}}$ and $z_{\textrm{s}}$ ranges, the fraction of outliers is lower in the weighted case, reaching a maximum of $\sim$0.20. When studying the bias, it reaches $\sim$-0.02 at the faintest magnitudes in the \textsc{BCNZ} case, while it oscillates very close to 0 as a function of $i_{\textrm{AB}}$ in the weighted case. The bias increases up to $\sim$0.02 and $\sim$-0.04 as a function of $z_{\textrm{b}}$ and $z_{\textrm{s}}$, respectively, in the \textsc{BCNZ}w photo-$z$ for the last bin, although it stays very close to 0 in all the other redshift range. At the last redshift bin, the \textsc{BCNZ} photo-$z$ shows a lower bias than \textsc{BCNZ}w. Finally, the performance of the \textsc{BPZ} photo-$z$ is worse than the \textsc{BCNZ} photo-$z$ estimates in almost all the metrics analyzed.

\subsection{Colour separation}
\label{sec:colour separation}

One of the most relevant properties of galaxies is its morphological type, which is broadly divided into elliptical and spiral galaxies. Elliptical galaxies tend to be massive and dominated by old stellar populations, while spiral galaxies are less massive and more gas rich, allowing star formation to happen (\citealt{Hubble, Driver}). The morphology of galaxies is related to their colour, so that elliptical galaxies tend to be redder and spiral galaxies bluer. Colour can therefore be used as a proxy for morphology. 

\begin{figure*}
    \centering
    \includegraphics[width=0.95\textwidth]{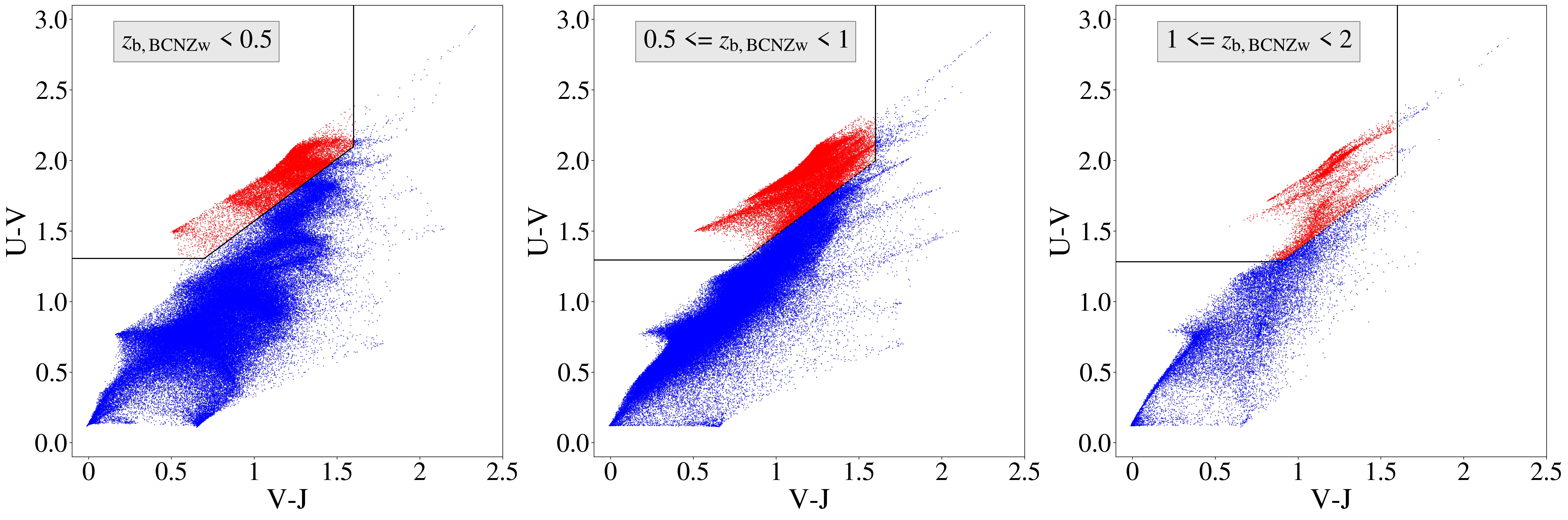}
    \caption{U-V vs. V-J planes used to define blue and red galaxies. The diagonal cut varies across different photo-$z$ ranges, shifting towards higher U-V values as the photo-$z$ increase.}
    \label{fig:colours_W1}
\end{figure*}

Studying the dependence of the photo-$z$ performance with respect to colour is useful since it is then possible to study science cases based on the morphology of the galaxies. To compute the rest-frame colour of a galaxy, we used CIGALE (\citealt{CIGALE}) to calculate the U-V and V-J colours (Siudek at al., in prep.). In Fig.~\ref{fig:colours_W1}, we illustrate the cuts we made in this colour-colour plane depending on the redshift of the galaxy (eq.~\ref{eq:colour_separation}), as was implemented in \cite{colourseparation}, where they studied star-forming galaxies up to $z \sim 2$:

\begin{equation}\label{eq:colour_separation}
    U-V \geq 0.88 \cdot V-J + \alpha,
\end{equation}
such that $\alpha=0.69$ for galaxies with $z_{\textrm{b, \textsc{BCNZ}w}}<0.5$, $\alpha=0.59$ for $0.5 \geq z_{\textrm{b, \textsc{BCNZ}w}}<1$ and $\alpha=0.49$ for $z_{\textrm{b, \textsc{BCNZ}w}}\geq 1$, where $z_{\textrm{b, \textsc{BCNZ}w}}$ corresponds to the weighted photo-$z$ of the galaxy, which is used to better divide the sample in colour. Additional conditions $U-V>1.3$ and $V-J<1.6$ were implemented in order to avoid contamination from unobscured and dusty star-forming galaxies (\citealt{colourseparation}). All galaxies fulfilling these conditions are defined as red galaxies, while the rest of them are considered as blue. Using this definition, we obtain 309504 red galaxies and 1240605 blue ones. Physical properties derived through CIGALE have been already used in PAUS to distinguish between red and blue galaxies (\citealt{Tortorelli, IA_Johnston}) as well as to validate the recreation of spectral features (\citealt{D4000}).

\begin{figure*}
    \centering
    \includegraphics[width=0.95\textwidth]{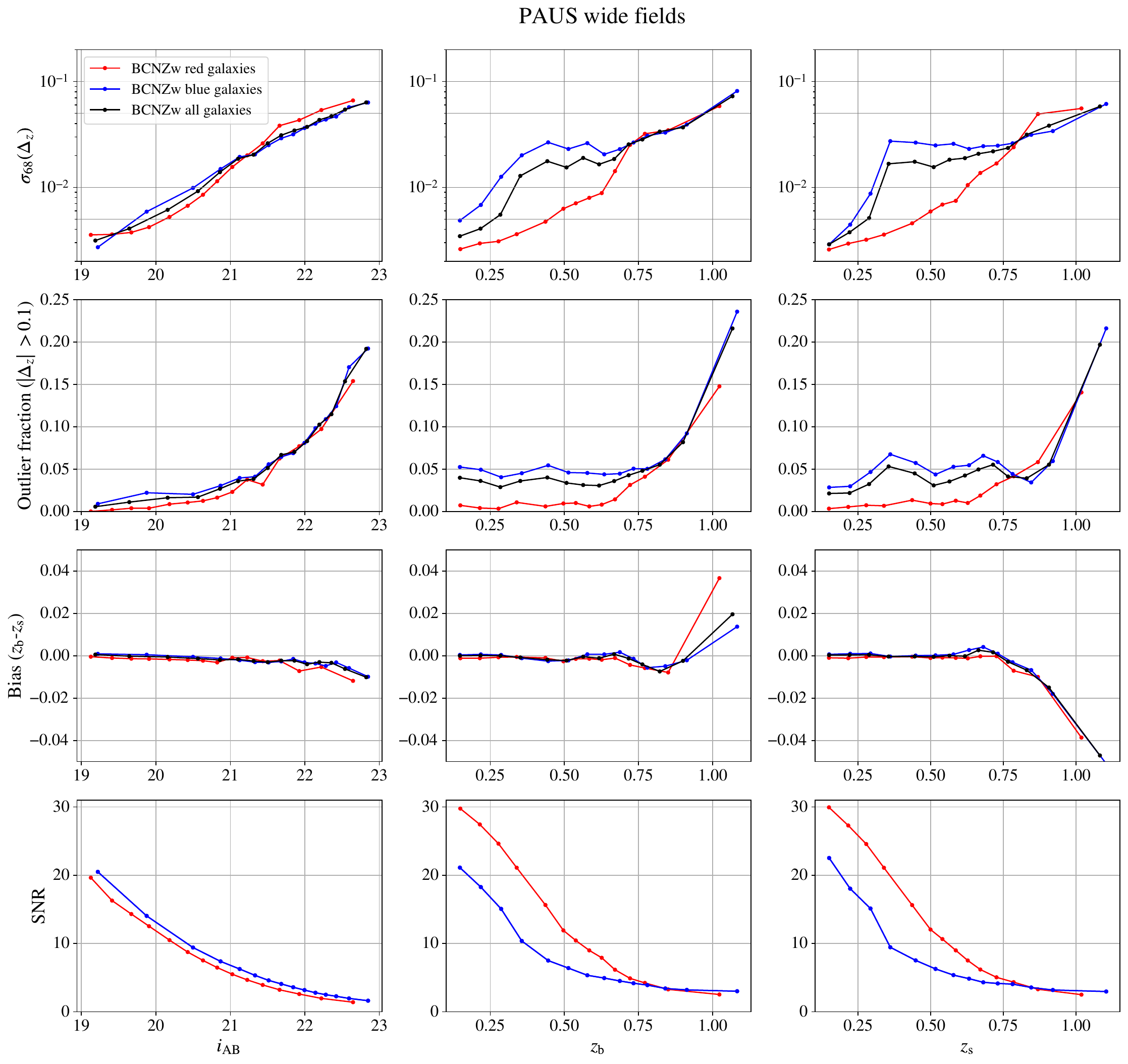}
    \caption{Performance of the PAUS wide fields for red, blue and all galaxies. From top to bottom, the $\sigma_{68}$, the outlier fraction, the bias and the SNR are shown as a function of the $i_{\textrm{AB}}$, the photometric and the spectroscopic redshift. Each bin is defined to contain an equal number of objects. Bright and low redshift red galaxies exhibit superior performance compared to their blue counterparts. Conversely, faint and high redshift blue galaxies demonstrate better performance.}
    \label{fig:metrics_all_WFs_colour}
\end{figure*}

Fig.~\ref{fig:metrics_all_WFs_colour} shows, from top to bottom, the $\sigma_{68}$, the outlier fraction, the bias and the SNR as a function of the $i_{\textrm{AB}}$, the photometric redshift and the spectroscopic redshift (from left to right) for red, blue and all galaxies for the PAUS wide fields. We include the SNR in this analysis to better understand the photo-$z$ performance as a function of $z_{\textrm{b}}$ and $z_{\textrm{s}}$. The percentage of red galaxies oscillates between $\sim$$20\%-30\%$ in all the redshift bins and between $\sim$$50\%-10\%$ in the magnitude bins, being lower the percentage of red galaxies for faint and high redshift objects. The panel with $\sigma_{68}$ as a function of $i_{\textrm{AB}}$ shows that red galaxies have better photo-$z$ statistics than blue galaxies for bright objects, while for faint objects, red galaxies perform worse, showing larger photo-$z$ scatter. For low redshift galaxies, the performance is also better for red objects, while it is slightly worse for those at higher redshifts. The outlier fraction for red galaxies is lower for almost all the $i_{\textrm{AB}}$ range while, as a function of the redshift, the outlier fraction for red galaxies is in general lower for the whole redshift range, specially at $z_{\textrm{b}}<0.75$. Regarding the bias, red galaxies present larger values than blue galaxies for faint and high redshift galaxies, while for the rest of the $i_{\textrm{AB}}$ and redshift range the bias is comparable between both types of galaxies. Finally, the SNR shows a clear difference between red and blue galaxies as a function of $z_{\textrm{b}}$ and $z_{\textrm{s}}$, with blue galaxies having lower SNR at low redshift bins. This can be explained by the fact that, for a given redshift bin, red galaxies are more massive and luminous than their blue counterparts, leading to higher SNR values and better photo-$z$ accuracy. In order to test the dependence on redshift without the effect of the SNR, we divided the sample in 3 bins of SNR: SNR < 3, 3<SNR<6, 6<SNR<12. We only found better performance for red galaxies in the highest SNR bin for all $z$, while in the other bins the performance was very similar and independent of $z$.

\subsection{Validation of the \texorpdfstring{$p(z)$}{z}}
\label{sec:Performance of the $p(z)$}

\textsc{BCNZ} provides us with the redshift probability distribution, $p(z)$, for each galaxy, where the photo-$z$ is determined by the maximum of that probability. However, a correct estimation of $p(z)$ allows us to use the photometric redshift not just as a point determined by the maximum, but as a probability density across the whole redshift space. This is useful for some science applications, such as \cite{Myers_pdf} and \cite{Asorey_pdf}, where they estimate galaxy clustering from the full $p(z)$ distribution. Nevertheless, the probability distribution obtained from SED template-fitting codes may be biased by degeneracies in the colour-redshift relation, non-representative SED templates, focusing effects and other inaccuracies (\citealt{PIT_euclid}). Thus, in this section we try to correct the $p(z)$ that \textsc{BCNZ} outputs and we compare its performance with the photo-$z$ given by the maximum of the $p(z)$.

The usual way of determining the validity of the $p(z)$ is by studying the probability integral transform (PIT, \citealt{PIT}), which accounts for the cumulative probability distribution up to the spectroscopic redshift of a given galaxy, defined as:
\begin{equation}
    \zeta(z_{\textrm{s}})=\int_{0}^{z_{\textrm{s}}} p(z) {\rm d}z,
\end{equation}
where the integral of the redshift probability distribution is carried out from 0 to the spectroscopic redshift ($z_{\textrm{s}}$).

For accurate $p(z)$, it is expected that the PIT values ($\mathrm{N_{PIT}}$) will be uniformly distributed over the range from 0 to 1. However, for badly estimated $p(z)$, the $\mathrm{N_{PIT}}$ will be biased. If the $p(z)$ of the galaxies are too narrow in comparison to the distribution of the spectroscopic redshifts, $\mathrm{N_{PIT}}$ will be concave. In the opposite case, the $p(z)$ will be too broad and $\mathrm{N_{PIT}}$ will be convex (\citealt{Polsterer}). It is possible to calibrate the $p(z)$ and obtain a uniform distribution of $\mathrm{N_{PIT}}$ by following the procedure in \cite{Bordoloi}, where they correct the biased $p(z)$ by multiplying it by the cumulative probability distribution at each $z$, obtaining a corrected $p(z)$ ($p_{\textrm{corr}}(z)$):
\begin{equation}\label{eq:pz_correction}
    p_{\textrm{corr}}(z) = p(z) \cdot N_{PIT}(\zeta(z)).
\end{equation}

Even though this correction of $p(z)$ is only strictly valid for the subsample with spec-$z$ information, it is reasonable to assume that applying it to the whole sample will still be valid if the spectroscopic sample is representative of the full catalogue.

\begin{figure}
    \centering
    \includegraphics[width=0.45\textwidth]{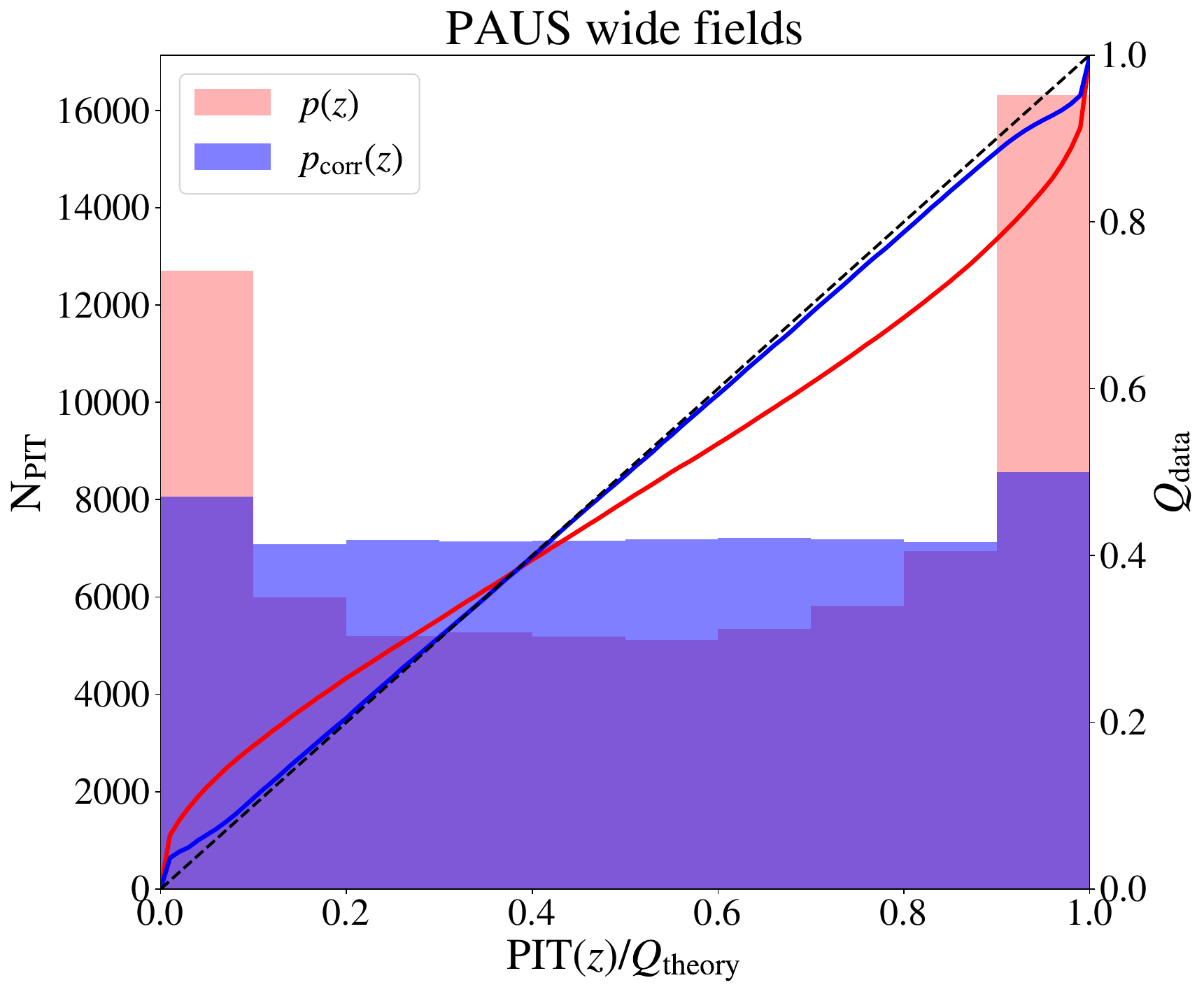}
    \caption{Distribution of the PIT values computed for the $p(z)$ before (red) and after (blue) being corrected, following \citealt{Bordoloi}, for the PAUS wide fields. The Quantile-Quantile plots, showing the fraction of spectroscopic redshift below a certain PIT value, are also shown. Before correcting the $p(z)$, the distribution of the PIT values show peaks at the edges, while the distribution becomes uniform after correcting it, as expected. The Quantile-Quantile plots also follow the diagonal line after the correction.}
    \label{fig:PIT_all}
\end{figure}

Fig.~\ref{fig:PIT_all} shows the $\mathrm{N_{PIT}}$ obtained from the $p(z)$ before and after being corrected for the PAUS wide fields. This figure also shows the Quantile-Quantile (QQ) plot, representing for each PIT value the fraction of spectroscopic redshifts found below it. The black dashed line shows the ideal case. We can see that the new $\mathrm{N_{PIT}}$, computed from the corrected $p(z)$, are now uniformly distributed and the QQ plots are much closer to the diagonal. It is also noticeable that, before being corrected, the $\mathrm{N_{PIT}}$ presents peaks at the edges of the PIT values, indicating that the uncorrected $p(z)$ are narrower than the distribution of the spectroscopic redshifts.

\begin{figure}
    \centering
    \includegraphics[width=0.45\textwidth]{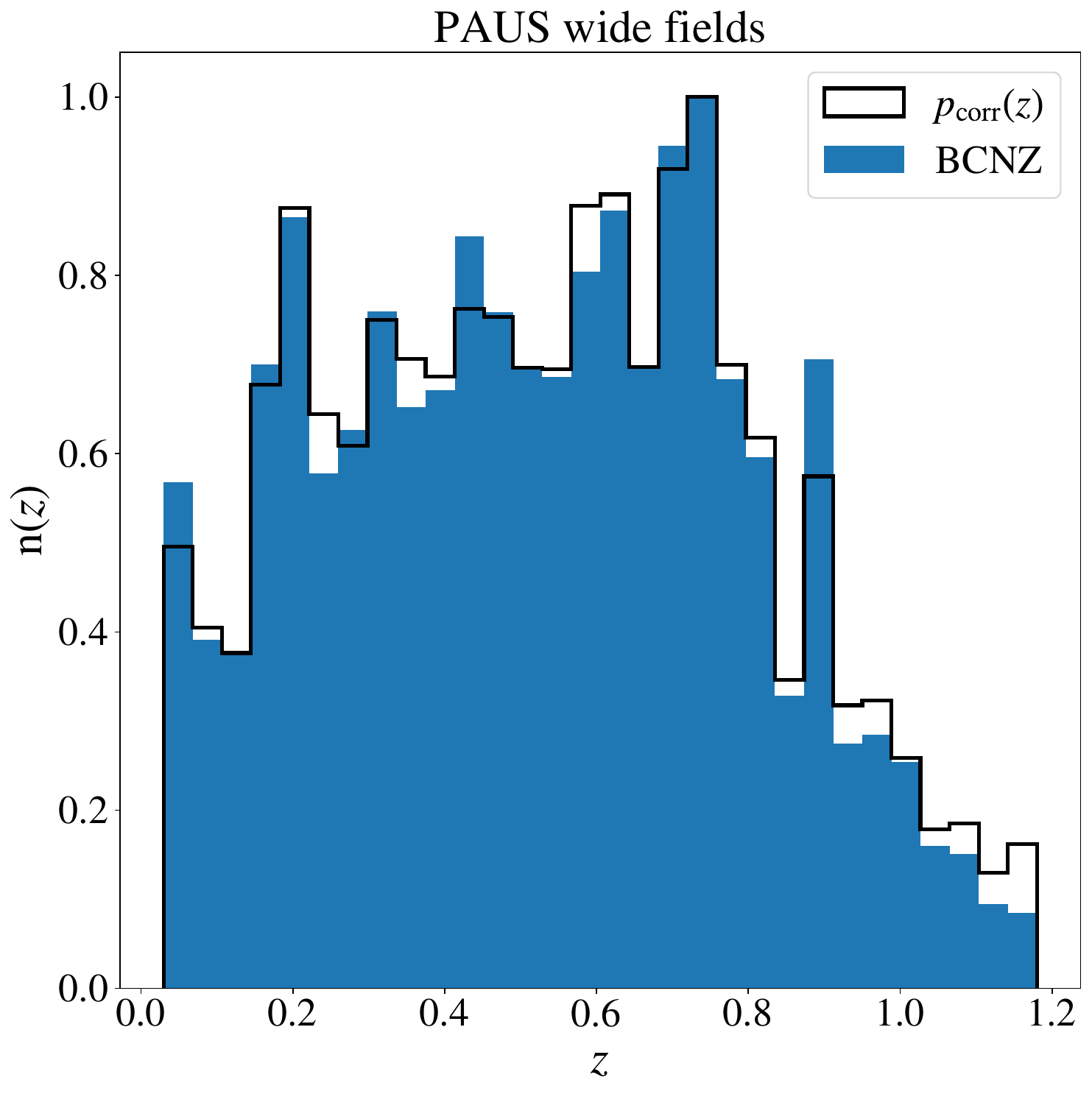}
    \caption{Normalized n($z$) for the PAUS wide fields as a function of the photometric redshift computed using the corrected $p(z)$ (black lines) and the estimation from the maximum of the distribution (\textsc{BCNZ}, blue histogram).}
    \label{fig:nz_pz_all}
\end{figure}

Fig.~\ref{fig:nz_pz_all} shows the normalized n($z$) distributions of the maximum of $p(z)$ (blue histogram), which corresponds to the $z_{\textrm{b, \textsc{BCNZ}}}$ estimate, and the sum of all the individual $p_{\textrm{corr}}(z)$ (black line). We note the change in the distribution when using a different estimate of the photo-$z$. In the case of comparing the performance of the photo-$z$ from the maximum of the distribution or from the whole distribution, we only consider objects with photo-$z$ < 1.2, since some edge effects were being obtained if the analysis was extended until $z=2$. We believe that the reason for this to happen is that the number of objects in the validation sample is so low in the range of $z=[1.2, 2]$ that the correction performed in eq.~\ref{eq:pz_correction} fails.

\begin{figure*}
    \centering
    \includegraphics[width=0.95\textwidth]{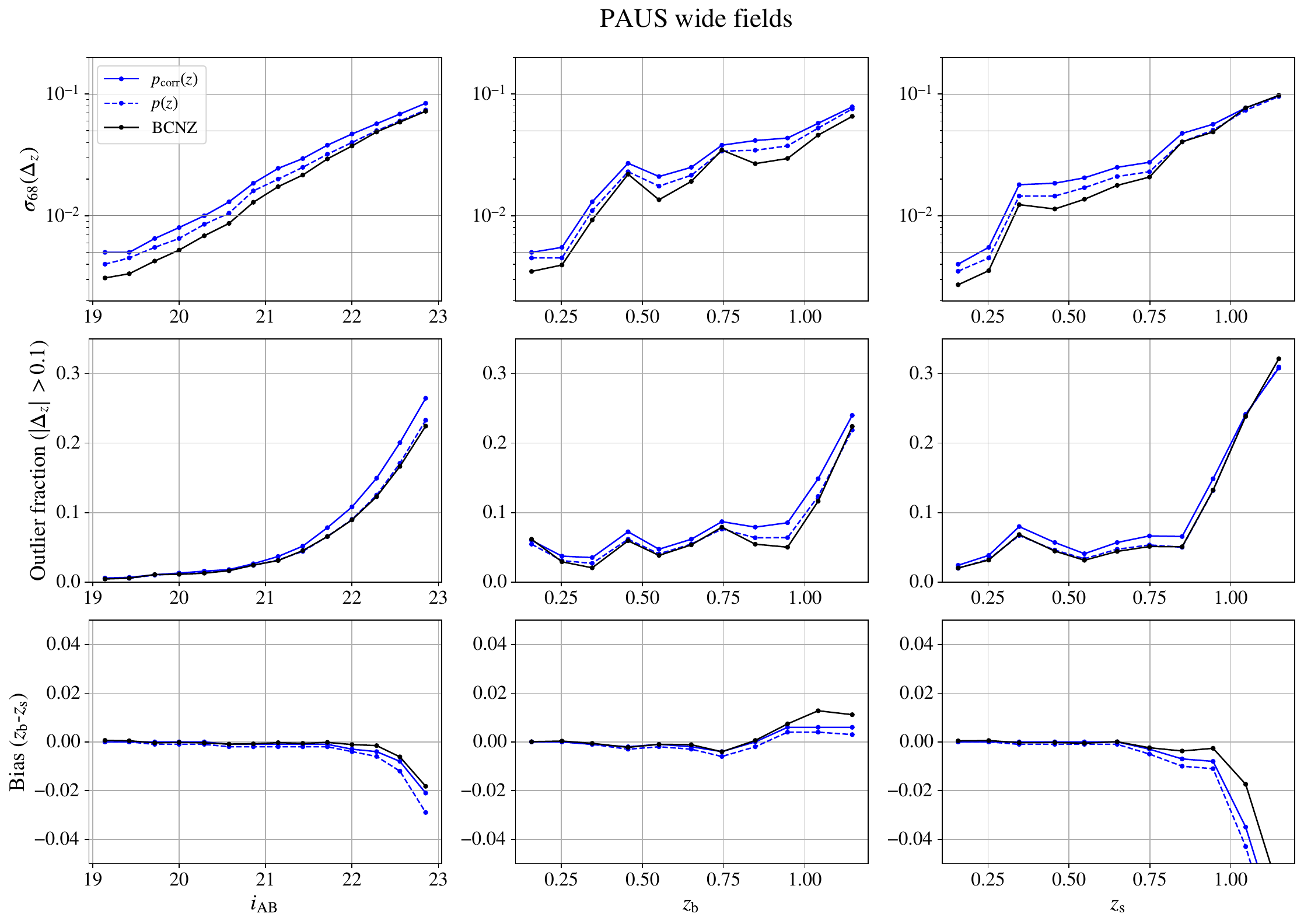}
    \caption{Performance of the PAUS wide fields from the maximum of the $p(z)$ (\textsc{BCNZ}), from $p(z)$ and from the corrected $p(z)$ ($p_{\textrm{corr}}(z)$). From top to bottom, the $\sigma_{68}$, the outlier fraction and the bias are shown as a function of the $i_{\textrm{AB}}$, the photometric and the spectroscopic redshift. Each bin is defined to contain an equal number of objects. Estimating the $z_{\textrm{b}}$ from the maximum of the $p(z)$ distribution demonstrates better performance than using the full $p_{\textrm{corr}}(z)$.}
    \label{fig:metrics_3_WFs_pz}
\end{figure*}

Fig.~\ref{fig:metrics_3_WFs_pz} shows the performance of $p(z)$ and $p_\textrm{corr}(z)$ (blue lines) in comparison with the one obtained from the maximum of $p(z)$ (names as \textsc{BCNZ}, black lines). Again, from top to bottom, we study the $\sigma_{68}$, the outlier fraction and the bias as a function of $i_{\textrm{AB}}$, $z_{\textrm{b}}$ and $z_{\textrm{s}}$, from left to right, respectively. In this case, the metrics for $p(z)$ and $p_\textrm{corr}(z)$ have been computed differently than before, since we do not define $\Delta_z$ (eq.~\ref{eq:dx}) from the maximum of the probability distribution, but for each $z$ in $p(z)$ and $p_\textrm{corr}(z)$. As a consequence, we can transform $p_\textrm{corr}(z)$ ($p(z)$) to $p_\textrm{corr}(\Delta_{z})$ ($p(\Delta_{z})$), with $\Delta_{z}=0$ corresponding to the $z_{\textrm{s}}$ of each object. Then, we stack the $p_\textrm{corr}(\Delta_{z})$ ($p(\Delta_{z})$) of all the objects, $i$, in each bin under study and obtain $\Sigma_{i} p_{i, \textrm{corr}}(\Delta_{z})$ ($\Sigma_{i} p_{i}(\Delta_{z})$), from where we compute the desired metrics. From the $\sigma_{68}$ plots, we see that $p_\textrm{corr}(z)$ presents higher values than $p(z)$. The reason for this is that the correction applied to $p(z)$ may distribute the probability away from its maximum, giving a more realistic $p_\textrm{corr}(z)$. Given that the $\sigma_{68}$ measures the width around $z_{\textrm{s}}$, it is expected that this width will be larger when distributing the probability away. In the case of the maximum of $p(z)$, the $\sigma_{68}$ presents lower values as a function of all the variables, with the exception of the faint and high redshift bins, where the $\sigma_{68}$ of $p(z)$ and \textsc{BCNZ} are very similar. In the case of the outlier fraction, the maximum of $p(z)$ also presents lower values than $p_\textrm{corr}(z)$. As for the bias, it is more pronounced as a function of $i_{\textrm{AB}}$ and $z_{\textrm{s}}$ for $p(z)$, while the \textsc{BCNZ} and $p_\textrm{corr}(z)$ cases are more similar in all the variables, with the bias of $p_\textrm{corr}(z)$ being closer to 0 as a function of $z_{\textrm{b}}$. As a conclusion, this analysis shows that the photo-$z$ accuracy reached by using the maximum of the $p(z)$ is higher than using the full $p_\textrm{corr}(z)$ distribution.

\section{Conclusions}
\label{sec:Conclusions}

The Physics of the Accelerating Universe Survey (PAUS) is based on a unique camera that uses a 40 NB set-up, allowing the fluxes of many objects to be obtained at once and their SEDs to be reconstructed with much more precision than typical BB surveys. The PAUS wide fields coverage overlap with the W1 and W3 fields from CFHTLenS and the KiDS observations in the GAMA G09 field, covering a total area of $\sim$51 deg$^{2}$. Here, we show how the BB photometry from these fields can be homogenised to a degree that permits combining them. This allows us to present photometric redshifts for all the PAUS wide fields, for $\sim$1.8 million objects, down to $i_{\textrm{AB}} = 23$ in a photo-$z$ range of $0 < z_{\textrm{b}} < 2$, with the vast majority of objects having $0 < z_{\textrm{b}} < 1.2$.

We compute photometric redshifts with a SED template-based algorithm called \textsc{BCNZ}, introducing a modification in the zero-point calibration technique between observed and modelled fluxes that allows us to be independent from spectroscopic samples, without degrading the performance of the photo-$z$.

Since the direct output from \textsc{BCNZ} ($z_{\textrm{b, \textsc{BCNZ}}}$) degrades in performance for faint objects, where the NB photometry has low signal-to-noise, we introduce an innovative weighting scheme that combines \textsc{BCNZ} with the broad band photo-$z$ from \textsc{BPZ} ($z_{\textrm{b, \textsc{BCNZ}w}}$). This weighted photo-$z$ estimate not only yields better accuracy for faint objects ($i_{\textrm{AB}} > 22.5$), but also shows a reduced percentage of outliers.

The abundance of galaxies as a function of magnitude and the photo-$z$ performance in each individual PAUS wide field are comparable. This enables us to study the performance of the combined fields in the main text, leaving for Appendix~\ref{sec:Study fields} the analysis for the individual fields. We obtain a $\sigma_{68}$ of 0.003 for the brightest objects ($i_{\textrm{AB}}\sim19$) and a $\sigma_{68}$ of $\sim$0.09 and $\sim$0.06 for the faintest ($i_{\textrm{AB}}\sim23$) in the \textsc{BCNZ} and \textsc{BCNZ}w photo-$z$ estimates, respectively. Using the weighted photo-$z$ estimate introduced in this work (\textsc{BCNZ}w), we manage to reduce the outlier fraction when studied as a function of $i_{\textrm{AB}}$, $z_{\textrm{b}}$ and $z_{\textrm{s}}$.

We also study the dependence of the photo-$z$ performance on galaxy colours, with colours determined from rest-frame absolute magnitudes. In general, we find that for bright ($i_{\textrm{AB}}<21$), low redshift objects ($z_{\textrm{b}}$ and $z_{\textrm{s}}< 0.75$), red galaxies have more accurate photo-$z$. In contrast, for faint objects, blue galaxies have slightly better photo-$z$ estimates. However, the photo-$z$ performance as a function of colour is relatively similar for both populations.

Finally, we validate and calibrate the probability density distribution, $p(z)$, given by \textsc{BCNZ}. We show how to properly correct $p(z)$ to follow the theoretical PIT distributions, indicating that they are well calibrated. This allows us to estimate the photo-$z$ not only as single-point measurement, but to assign a robust photo-$z$ probability to each object. Nevertheless, we find better photo-$z$ accuracy by taking the maximum of $p(z)$ as the photo-$z$ estimate rather than using the full distribution.

The photo-$z$ catalogues presented here are being used for key science projects in PAUS, in particular the measurement of galaxy intrinsic alignments and galaxy clustering as a function of redshift, magnitude and colour (Navarro-Gironés et al, in prep), in a rather unique regime of moderately faint and high redshift samples, extending the analysis to $i_{\textrm{AB}}\sim23$ and $z_{\textrm{b}}\sim1$. This regime is of utmost importance to calibrate weak lensing analysis from dark energy missions such as Euclid (\citealt{Euclid}) or Rubin LSST (\citealt{LSST}). These samples can also be used to calibrate the redshift distributions for such missions, as was already done with previous PAUS data in the case of KiDS (\citealt{KiDS_photoz_calibration}) and DES (\citealt{DES_photoz_calibration}). The photometric redshift catalogues presented in this paper are published under demand as part of the PAUS legacy products.

\section*{Acknowledgements}

DNG, EG and MC acknowledge support from the Spanish Ministerio de Ciencia e Innovacion (MICINN), project PID2021-128989NB. HH is supported by a DFG Heisenberg grant (Hi 1495/5-1), the DFG Collaborative Research Center SFB1491, as well as an ERC Consolidator Grant (No. 770935). AHW is supported by the Deutsches Zentrum für Luft- und Raumfahrt (DLR), made possible by the Bundesministerium für Wirtschaft und Klimaschutz, and acknowledges funding from the German Science Foundation DFG, via the Collaborative Research Center SFB1491 "Cosmic Interacting Matters - From Source to Signal". MS acknowledges the support by the Polish National Agency for Academic Exchange (Bekker grant BPN/BEK/2021/1/00298/DEC/1), the European Union's Horizon 2020 Research and Innovation programme under the Maria Sklodowska-Curie grant agreement (No. 754510). PR acknowledges the support by the Tsinghua Shui Mu Scholarship, the funding of the National Key R\&D Program of China (grant no. 2018YFA0404503), the National Science Foundation of China (grant no. 12073014 and 12350410365), the science research grants from the China Manned Space Project with No. CMS-CSST2021-A05, and the Tsinghua University Initiative Scientific Research Program (No. 20223080023) EJG acknowledges support from the Spanish Research Project PID2021-123012NA-C44. CMB acknowledges support from the Science Technology Facilities Council (ST/T000244/1, ST/X001075/1). GM acknowledges support from the Collaborative Research Fund under Grant No. C6017-20G which is issued by the Research Grants Council of Hong Kong S.A.R. JC acknowledges support from the grant PID2021-123012NA-C44 funded by MCIN/AEI/ 10.13039/501100011033 and by “ERDF A way of making Europe”.

The PAU Survey is partially supported by MINECO under grants CSD2007-00060, AYA2015-71825, ESP2017-89838, PGC2018-094773, PGC2018-102021, PID2019-111317GB, SEV-2016-0588, SEV-2016-0597, MDM-2015-0509 and Juan de la Cierva fellowship and LACEGAL and EWC Marie Sklodowska-Curie grant No 734374 and no.776247 with ERDF funds from the EU Horizon 2020 Programme, some of which include ERDF funds from the European Union. IEEC and IFAE are partially funded by the CERCA and Beatriu de Pinos program of the Generalitat de Catalunya. Funding for PAUS has also been provided by Durham University (via the ERC StG DEGAS-259586), ETH Zurich, Leiden University (via ERC StG ADULT-279396 and Netherlands Organisation for Scientific Research (NWO) Vici grant 639.043.512), University College London and from the European Union's Horizon 2020 research and innovation programme under the grant agreement No 776247 EWC. The PAU data center is hosted by the Port d'Informaci\'o Cient\'ifica (PIC), maintained through a collaboration of CIEMAT and IFAE, with additional support from Universitat Aut\`onoma de Barcelona and ERDF. We acknowledge the PIC services department team for their support and fruitful discussions.

This work has made use of CosmoHub. CosmoHub has been developed by the Port d'Informació Científica (PIC), maintained through a collaboration of the Institut de Física d'Altes Energies (IFAE) and the Centro de Investigaciones Energéticas, Medioambientales y Tecnológicas (CIEMAT) and the Institute of Space Sciences (CSIC \& IEEC).
CosmoHub was partially funded by the "Plan Estatal de Investigación Científica y Técnica y de Innovación" program of the Spanish government, has been supported by the call for grants for Scientific and Technical Equipment 2021 of the State Program for Knowledge Generation and Scientific and Technological Strengthening of the R+D+i System, financed by MCIN/AEI/ 10.13039/501100011033 and the EU NextGeneration/PRTR (Hadoop Cluster for the comprehensive management of massive scientific data, reference EQC2021-007479-P) and by MICIIN with funding from European Union NextGenerationEU(PRTR-C17.I1) and by Generalitat de Catalunya.

\section*{Data Availability}

The PAUS wide field catalogues will be shared on reasonable request to the corresponding authors. They can be accessed through \href{https://cosmohub.pic.es/home}{CosmoHub} (\citealt{CosmoHub1, CosmoHub2}).





\appendix

\section{Photo-\texorpdfstring{$z$}{z} performance in individual fields}
\label{sec:Study fields}

In this Appendix, we analyse the performance of the photo-$z$ studied in Section~\ref{sec:Comparison Validation}, but stressing the comparison between each of the wide fields.

Table~\ref{tab:Photo-z_fields} shows the number of photo-$z$ (after applying the mask and the star flags), the masked area, the number density, the \textsc{BCNZ} $\sigma_{68}$ and the \textsc{BCNZ}w $\sigma_{68}$ for the PAUS wide fields studied in this analysis. The G09 number density is slightly smaller than that of the W1 and W3 fields. Nonetheless, the number densities are comparable, indicating that a similar population was selected in the three PAUS wide fields, as shown in Fig.~\ref{fig:number_counts}. As for both $\sigma_{68}$ estimates, the values for the W1 and W3 fields are almost the same, while in the G09 field this value increases up to $\sigma_{68}=0.026$, showing a lower performance.

\begin{table}
\caption{Number of photometric redshifts after applying the mask and rejecting stars, masked area, number density, \textsc{BCNZ} $\sigma_{68}$ and \textsc{BCNZ}w $\sigma_{68}$ for the W1, W3 and G09 fields observed by PAUS.}
\begin{center}
\label{tab:Photo-z_fields}
\begin{tabular}{c c c c}
\hline
\hline
Field & W1 & G09 & W3 \\ \hline
\# masked objects    &  308403 & 364592  & 568052  \\ \hline
Masked area [deg$^{2}$] & 10.2 & 12.54  & 18.25 \\ \hline
Number density [deg$^{-2}$] &  30236  & 29074 & 31126 \\ \hline
\textsc{BCNZ} $\sigma_{68}$  &  0.018  & 0.026  & 0.019 \\ \hline
\textsc{BCNZ}w $\sigma_{68}$  &  0.020  & 0.026  & 0.018 \\ \hline
\end{tabular}
\end{center}
\end{table}

\begin{figure*}
    \centering
    \includegraphics[width=1.\textwidth]{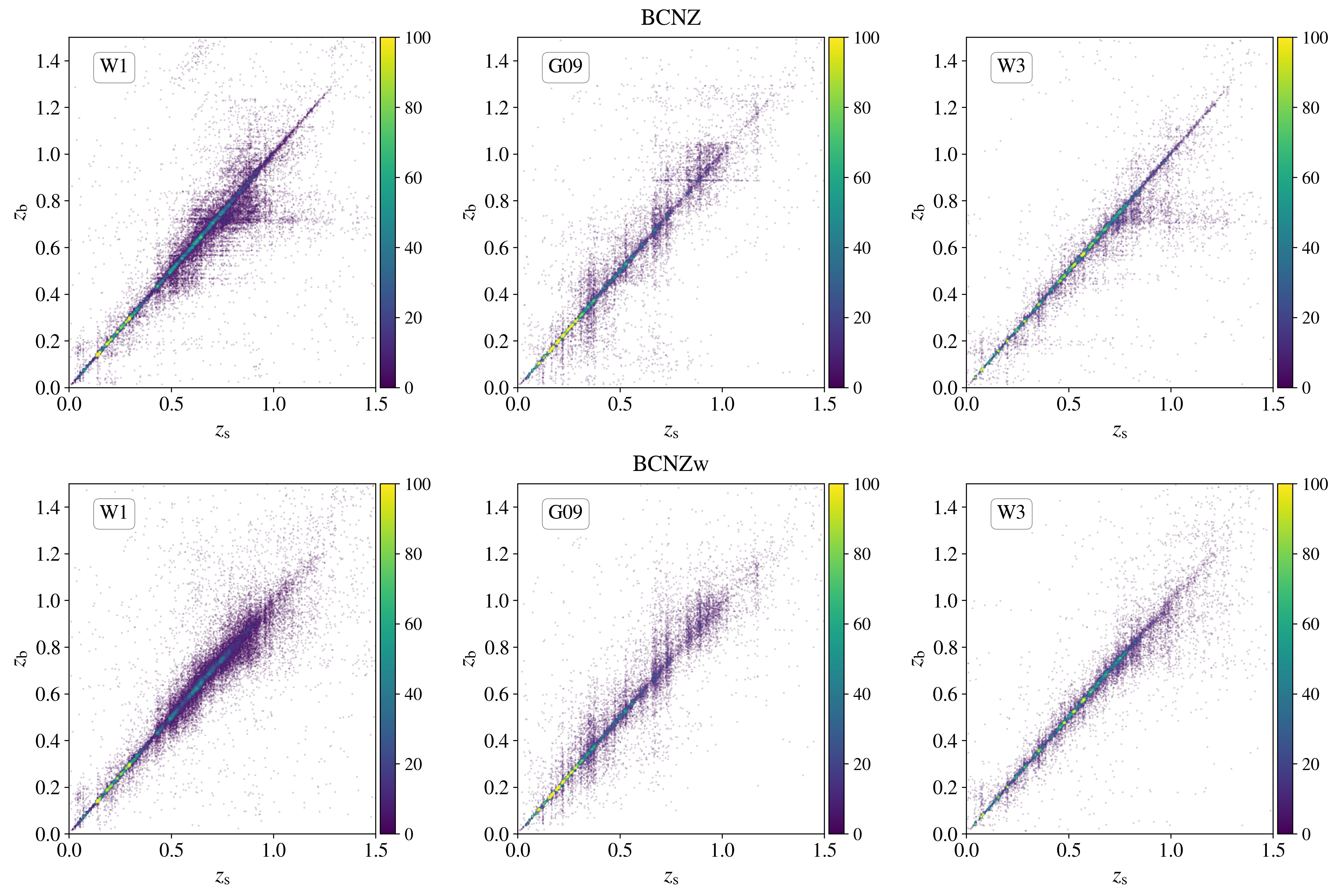}
    \caption{Photometric redshift vs spectroscopic redshift for the \textsc{BCNZ} photo-$z$ (top) and for the \textsc{BCNZ}w photo-$z$ (bottom) for each of the 3 PAUS wide fields. The colour bar indicates the density of objects. The horizontal stripes at $z_{\textrm{b}} \approx 0.72$ and $z_{\textrm{b}} \approx 0.89$ are dissipated when weighting with the photo-$z$ computed only with broad bands, which are obtained with another photometric redshift code.}
    \label{fig:weighted_zb_vs_zs_fields}
\end{figure*}

Fig.~\ref{fig:weighted_zb_vs_zs_fields} shows the photo-$z$ as a function of spec-$z$, as in the case of Fig.~\ref{fig:weighted_zb_vs_zs}, but separated into the 3 PAUS wide fields, so that we can see the effect of applying the weight on the photo-$z$ in each of the fields. A horizontal stripe of artificial photometric redshifts is seen at $z_{\textrm{b}} \sim$0.72 for the W1 and W3 fields and at $z_{\textrm{b}} \sim$0.89 for the G09 field, which is mainly corrected with the new weighted photo-$z$.

\begin{figure}
    \centering
    \includegraphics[width=0.46\textwidth]{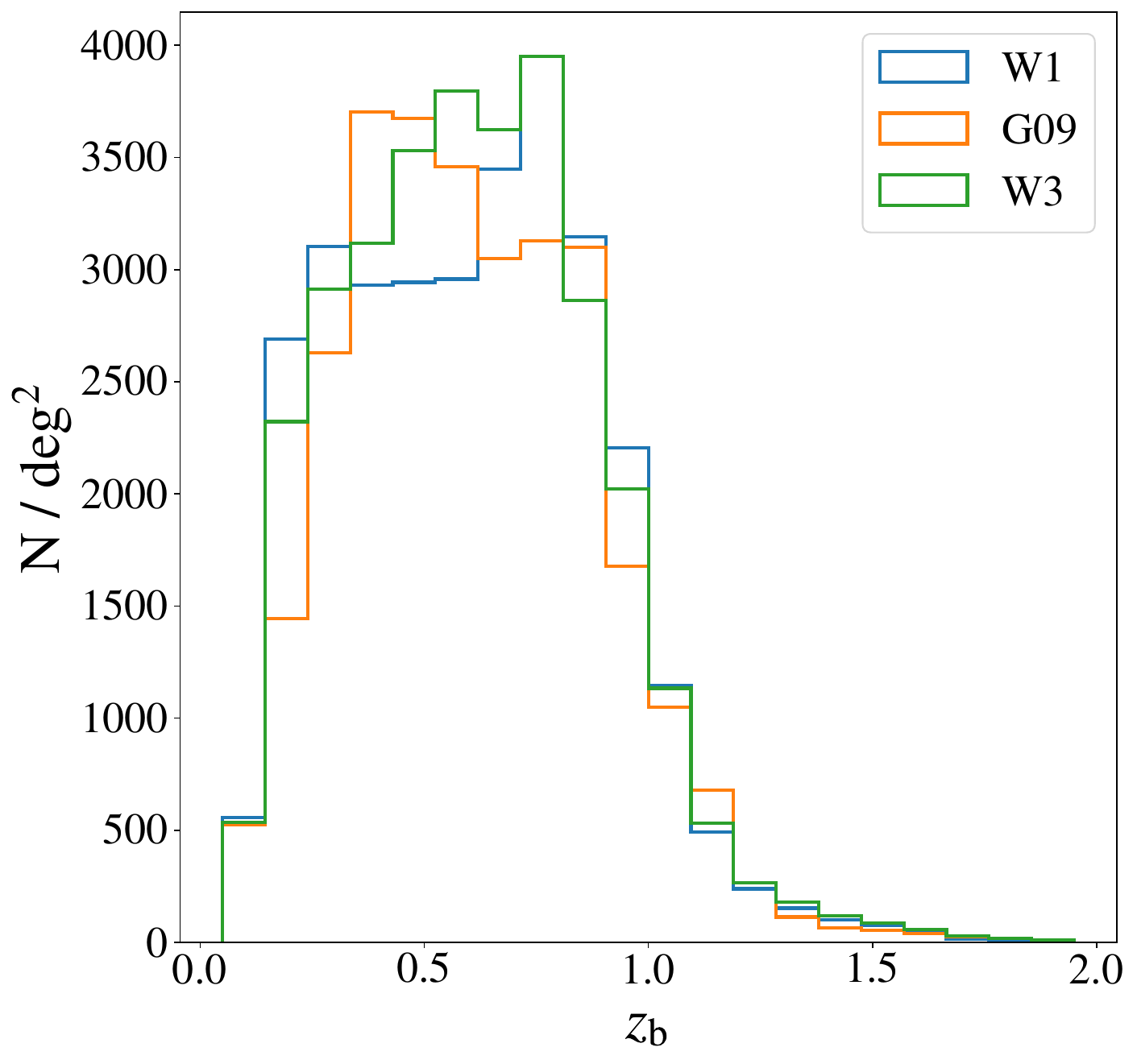}
    \caption{Weighted photometric redshift distributions for the W1, G09 and W3 fields, using the \textsc{BCNZ}w photo-$z$. Under and overdensities are observed between the PAUS wide fields.}
    \label{fig:nz_wide_fields}
\end{figure}

The weighted photometric redshift distributions are presented in Fig. \ref{fig:nz_wide_fields} for the W1, W3 and G09 fields. The W1 and W3 distributions are very similar, with the exception of an increase in the W3 distribution from $z_{\textrm{b}} \sim0.4-0.6$, not present in W1. In the case of the G09 field, there is an underdensity of objects at $z_{\textrm{b}} \sim0.3-0.4$ and an overdensity at $z_{\textrm{b}} \sim0.4-0.5$. However, it is not straightforward to assess if these are intrinsic differences coming from the fields themselves or if they are caused by the fact that we are comparing two photometric systems, with different star flags, different masks and different flux errors. 

\begin{figure*}
    \centering
    \includegraphics[width=0.95\textwidth]{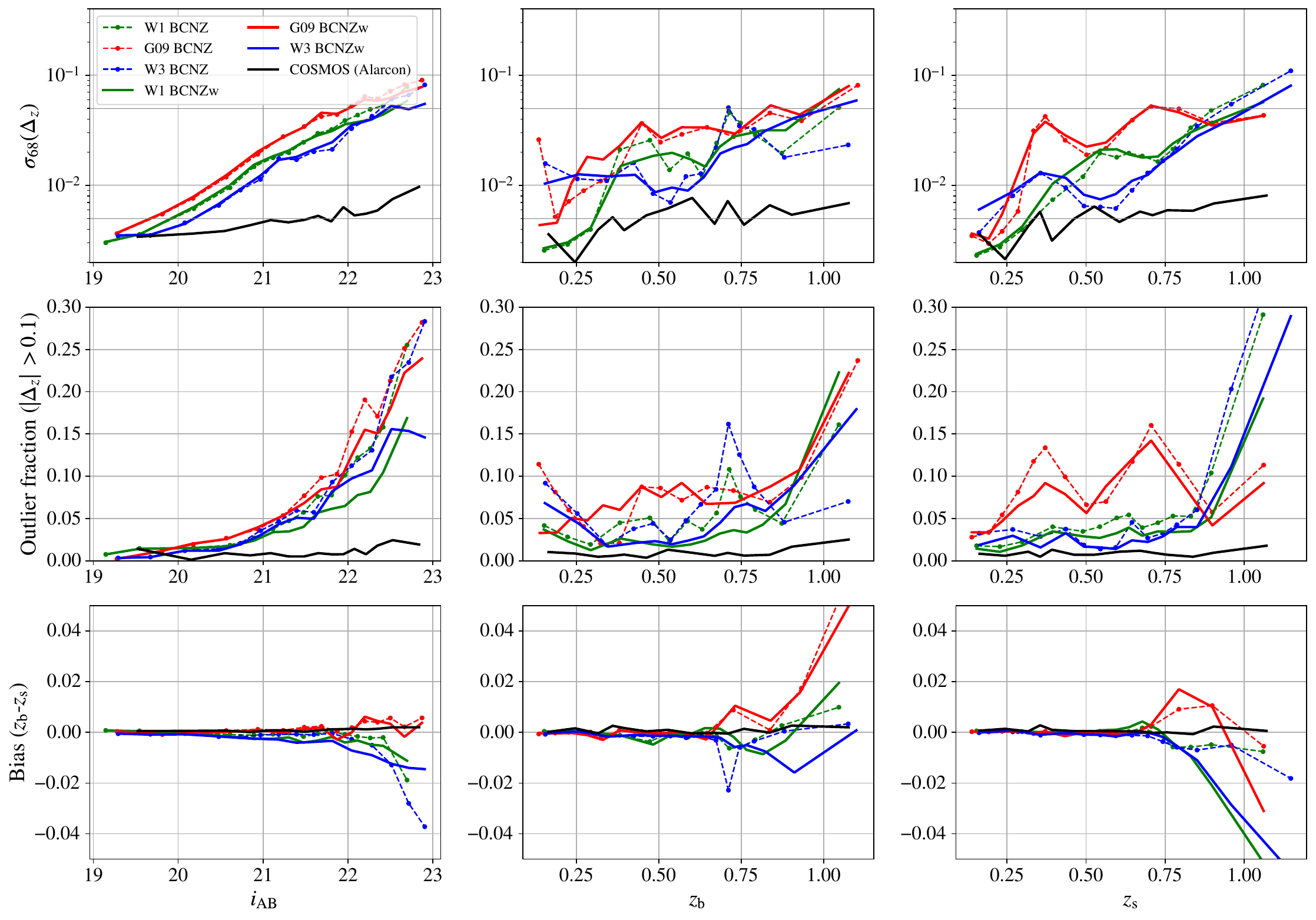}
    \caption{Performance of the W1, G09 and W3 fields (green, red and blue, respectively) for the \textsc{BCNZ}w photo-$z$ and for the COSMOS field photo-$z$ in \citealt{Alarcon_bcnz} (black). From top to bottom, the $\sigma_{68}$, the outlier fraction and the bias are shown as a function of $i_{\textrm{AB}}$ magnitude, photometric and spectroscopic redshift. Each bin is defined to contain an equal number of objects.}
    \label{fig:metrics_3_WFs}
\end{figure*}

Fig.~\ref{fig:metrics_3_WFs} shows the performance of the W1, W3 and G09 fields as a function of $i_{\textrm{AB}}$, $z_{\textrm{b}}$ and $z_{\textrm{s}}$ for the weighted photo-$z$ (solid lines) and the \textsc{BCNZ} photo-$z$ (dashed lines), following the same structure as Fig.~\ref{fig:metrics_all_WFs}. We also include in this analysis the photo-$z$ from the COSMOS field computed by \citealt{Alarcon_bcnz} (solid black lines), so as to compare the photo-$z$ with best accuracy that have been obtained by PAUS with the wide fields photo-$z$. Focusing on the PAUS wide fields, the $\sigma_{68}$ values as a function of $i_{\textrm{AB}}$ show that the performance of the G09 field is, in general, worse than the W1 and W3 fields, which are more comparable to one another. This is also observed when looking at the $\sigma_{68}$ as a function of $z_{\textrm{b}}$ and $z_{\textrm{s}}$, with the exception of the objects around $z_{\textrm{s}}=1$, where G09 performs slightly better. We note that the weighted photo-$z$, in general, present lower values of $\sigma_{68}$. For the \textsc{BCNZ}w photo-$z$, the faintest objects ($i_{\textrm{AB}}\sim22.5-23$) have a $\sigma_{68}\sim0.05-0.06$ for W1 and W3, while it arrives at $\sigma_{68}\sim0.08$ for G09. When studying the outlier fractions, we note that G09 presents higher values than the W1 and W3 cases, which are very similar. G09 reaches an outlier fraction of $\sim$0.25 for the faintest bin, while W1 and W3 present a value of $\sim$0.15. As a function of $z_{\textrm{b}}$ and $z_{\textrm{s}}$, the outlier fraction is higher in the G09 field at intermediate redshifts ($z\sim0.25-0.75$). As for the bias, we see that the behavior is different for the G09 case as a function of $i_{\textrm{AB}}$, showing positive values for the \textsc{BCNZ} photo-$z$, unlike the W1 and W3 fields, which show negative bias. G09 also presents higher bias as a function of $z_{\textrm{b}}$. Focusing now in the COSMOS photo-$z$ presented in \citealt{Alarcon_bcnz}, the $\sigma_{68}$ and the outlier fraction values are lower than in the PAUS wide fields, while the bias is comparable between the combined wide fields fields shown in Fig.~\ref{fig:metrics_all_WFs} and these COSMOS photo-$z$. It is important to highlight that the average number of PAUS observations in COSMOS is 5, while it is 3 in the PAUS wide fields. As a consequence, the SNR in COSMOS is higher than in the wide fields and the photo-$z$ are more accurate. Additionally, for these COSMOS photo-$z$, the number of bands used was higher, with 26 broad, intermediate and narrow bands covering the UV, visible and near IR, besides the 40 PAUS NB. Finally, we would like to mention that the number of objects and the area of the COSMOS photo-$z$ are much lower than the PAUS wide fields photo-$z$, with $\sim$40k versus $\sim$1.8 million objects and $\sim$1deg$^{2}$ versus $\sim$51deg$^{2}$, respectively. As a consequence, the PAUS wide fields photo-$z$ have better statistics and are less affected by sample variance.

\begin{figure*}
    \centering
    \includegraphics[width=0.85\textwidth]{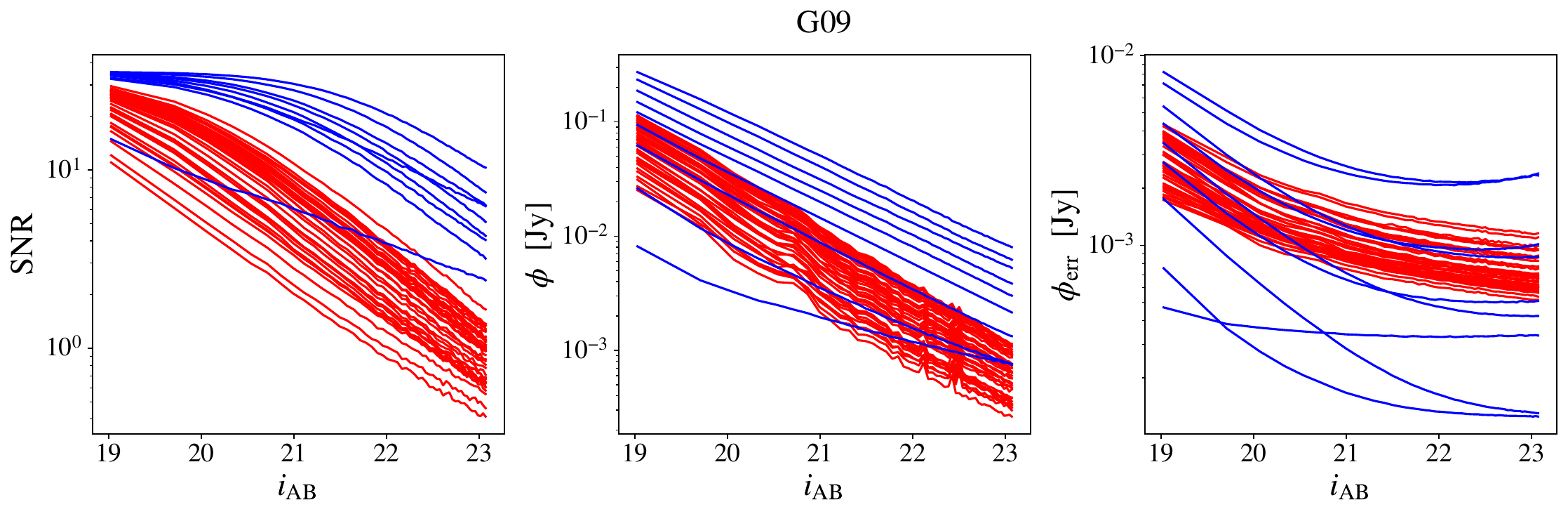}
    \caption{SNR, flux ($\phi$) and flux error ($\phi_{\mathrm{err}}$) (from left to right) of the W2 field for the broad bands (blue lines) and the narrow bands (red lines). The SNR of the broad bands starts to decrease at $i_{\textrm{AB}}\sim21$, since the flux errors of the G09 broad bands stop decreasing around the same magnitude. In the case of the narrow bands, the SNR decreases as the objects become fainter, which is caused by the constant NB flux errors at faint magnitudes.
    }
    \label{fig:SNR_weight_W2}
\end{figure*}

We now summarise the reasons why the G09 photo-$z$ display a poorer  performance than the other fields:
\begin{itemize}
    \item Fig.~\ref{fig:SNR_weight_W2} shows the SNR, flux ($\phi$) and flux error ($\phi_{\textrm{err}}$) of the G09 field. Note that the flux errors of the BB decrease at a slower rate than in the W3 field case shown in Fig.~\ref{fig:SNR_weight_W3} and stop decreasing at $i_{\textrm{AB}}\sim22$. As a consequence, the SNR of both NB and BB decrease for almost all the $i_{\textrm{AB}}$ range, with the SNR of the BB being still higher than the NB one. This lower BB SNR for the G09 field with respect to W1 and W3 is a key point that affects the photo-$z$ estimation, given that the SNR drives the photo-$z$ performance, as explained in Section~\ref{sec:Weighted photo-$z$}. However, it is important to note that this difference in SNR between G09 and W1/W3 may be magnified, since we expect the CFHTLenS errors to be underestimated in comparison with the KiDS ones. The reason for this is that KiDS uses GAaP photometry (\citealt{2008A&A...482.1053K}), where the convolution of the image is lower than in CFTHLenS and is corrected for in the photometric error estimates. Nevertheless, we still expect lower SNR in KiDS measurements, since CFHTLenS is deeper than KiDS in most of the broad bands.    
    \item Even though the KiDS photometric system has 9 broad bands instead of the 5 available in CFHTLenS, the VIKING near-infrared bands mostly help at $z>1$, where they can better detect the Balmer and 4000\r{A} break and where most of our objects are not located.
    \item The validation sample of the G09 field, which is obtained by merging the G09 and KiDZ-COSMOS spectroscopic redshifts, might not come from exactly the same processing, since the G09 objects are from KiDS DR4 and KiDZ-COSMOS come from KiDS DR5.
    \item The difference in the definition of the $i$-band between CFHTLenS and KiDS may also affect the comparison between both cases. On the one hand, even though we redefine $i_{\textrm{AB, CFHTLenS}} = i_{\textrm{AB, KiDS}} - 0.1$, we still have differences between both magnitudes, since this is only a first order correction, so that comparing the performance of KiDS and CFHTLenS as a function of $i_{\textrm{AB}}$ is not straightforward. On the other hand, as we commented, the cut at $i_{\textrm{AB, KiDS}}=23.1$ allows to obtain very similar number counts compared with W1 and W3. However, there are still some differences that can be observed for bright objects in Fig.~\ref{fig:number_counts} that may affect the comparison.
    \item Finally, the fact that different outliers are found in the W1/W3 and the G09 fields, which correspond to artificial photo-$z$ at $z_{\textrm{b}}\sim0.72$ and $z_{\textrm{b}}\sim0.89$, respectively, may affect the performance, even though we try to correct those outliers with the weighted photo-$z$.
\end{itemize}

\section{\texorpdfstring{$Q_{z}$}{Qz} separation}
\label{sec:Qz separation}

\textsc{BCNZ} provides photometric redshift quality parameters that allow us to select subsamples of galaxies with the ``best'' photo-$z$. As stated in Section~\ref{sec:Methodology}, we chose to use the parameter $Q_{z}$ (eq.~\ref{eq:Qz}), since it is a combination of other quality parameters.

\begin{figure*}
    \centering
    \includegraphics[width=0.95\textwidth]{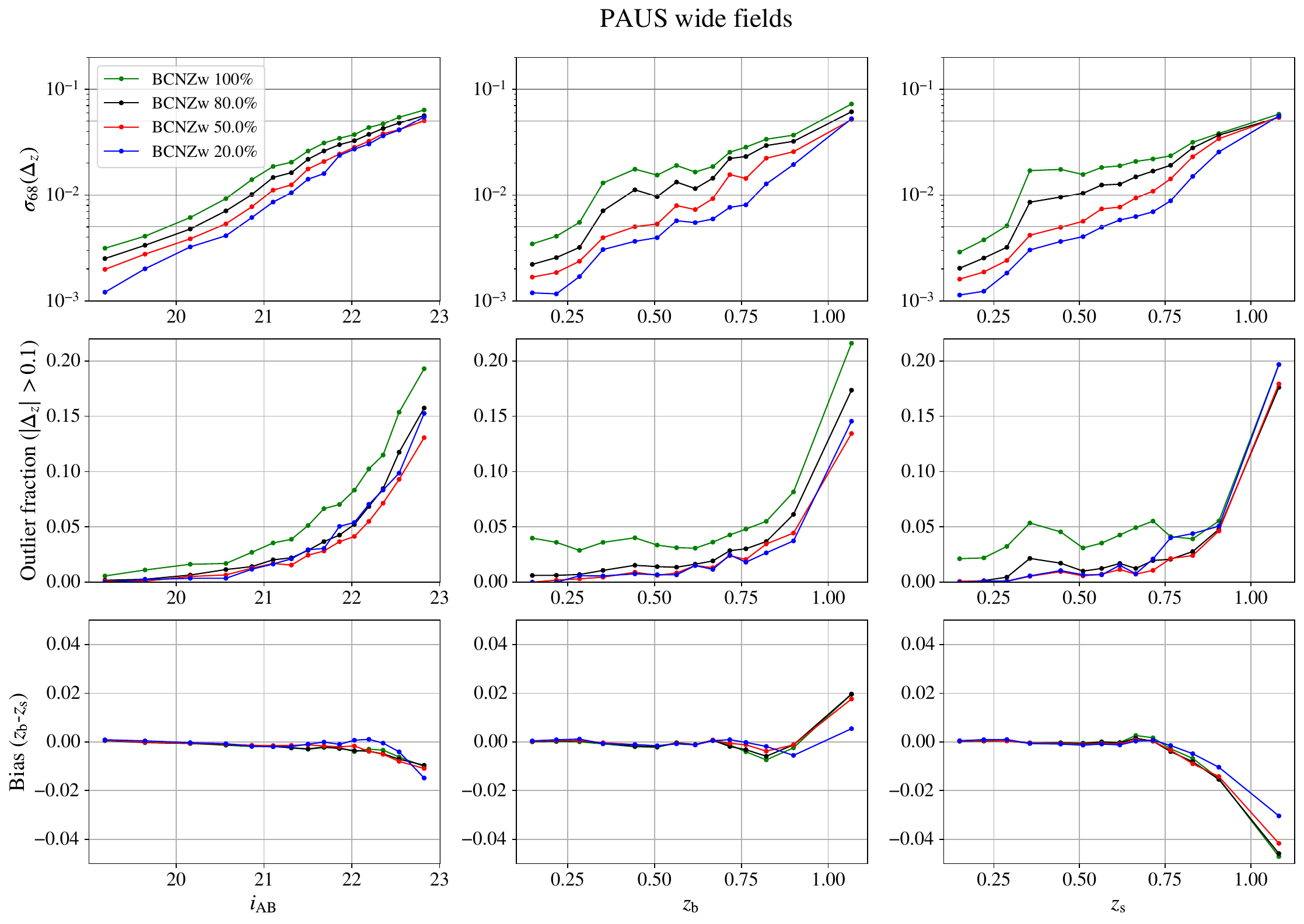}
    \caption{Performance of the PAUS wide fields as a function of the quality parameter $Q_{z}$ for the \textsc{BCNZ}w photo-$z$. From top to bottom, the $\sigma_{68}$, the outlier fraction and the bias are shown as a function of $i_{\textrm{AB}}$, the photometric and the spectroscopic redshift. The performance is shown for the 20, 50, 80 and 100\% best galaxies taking into account the quality parameter $Q_{z}$.}
    \label{fig:metrics_all_wfs_Qz}
\end{figure*}

\begin{figure*}
    \centering
    \includegraphics[width=0.95\textwidth]{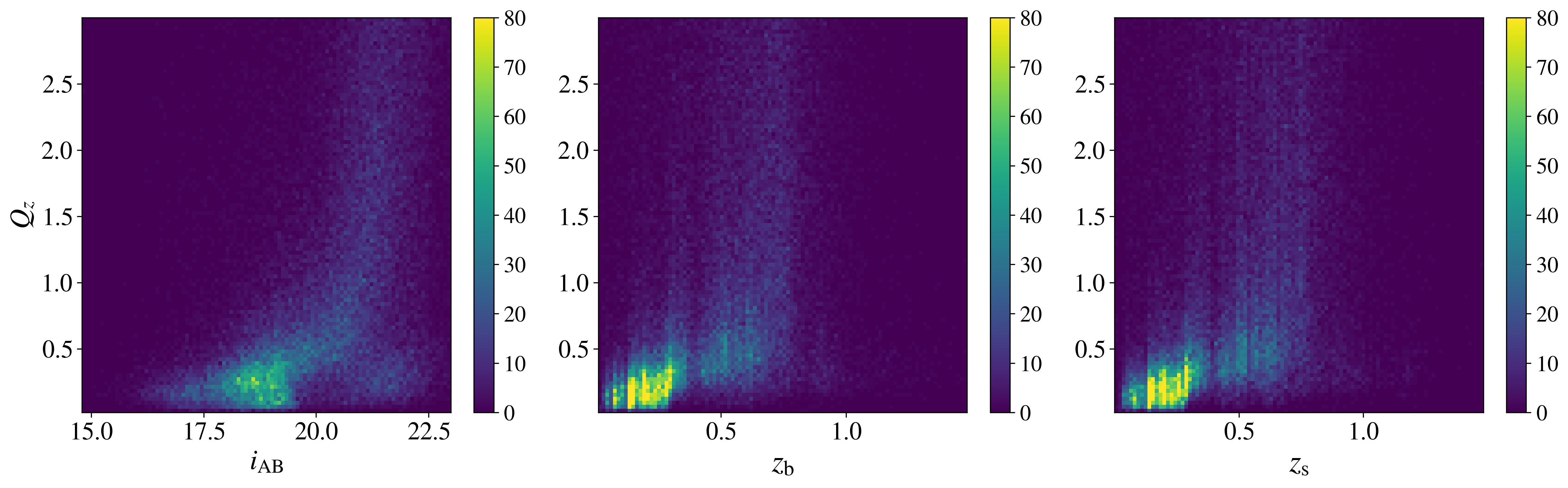}
    \caption{Dependence of the quality factor $Q_{z}$ on $i_{\textrm{AB}}$ (left), $z_{\textrm{b}}$ (middle) and $z_{\textrm{s}}$ (right). At high magnitudes ($i_{\textrm{AB}}>22$) and high redshifts ($z\sim1$), the values of $Q_{z}$ rise fast and are not that much correlated with the photo-$z$ accuracy.}
    \label{fig:Qz}
\end{figure*}

Fig.~\ref{fig:metrics_all_wfs_Qz} shows the performance as a function of the quality parameter $Q_{z}$ for the PAUS wide fields. Four different percentages of galaxies were selected based on $Q_{z}$, such that we retain 100\%, 80\%, 50\% or 20\% of objects in each of the bins under study with best photometric redshift estimates. Note that $Q_{z}$ acts reliably for objects until $i_{\textrm{AB}}<21.5$ and $z < 1$, since a more restrictive $Q_{z}$ cut in the catalogue yields better performance. However, for faint and high redshift objects, the performance does not improve much with the quality parameter cut. This can be seen in Fig.~\ref{fig:Qz}, where the dependence of $Q_{z}$ as a function of $i_{\textrm{AB}}$, $z_{\textrm{b}}$ and $z_{\textrm{s}}$ is shown.
Fig.~\ref{fig:metrics_all_wfs_Qz} can be useful in order to obtain a  catalogue with a better performance, taking into account the loss in the number of objects.

\section{Narrow band coverage}
\label{sec:NB coverage}

Due to the observing strategy in PAUS, some objects may lack measurements in some bands. Thus, the recovered SED may be less precise, since some emission lines could be lost and the general shape of the SED could be less defined. However, this effect might not be relevant if the number of bands with no measurements is low. Since \textsc{BCNZ} does not require flux measurements in all the bands to compute the photo-$z$, the performance can be studied as a function of the NB coverage.

\begin{figure*}
    \centering
    \includegraphics[width=0.4\textwidth]{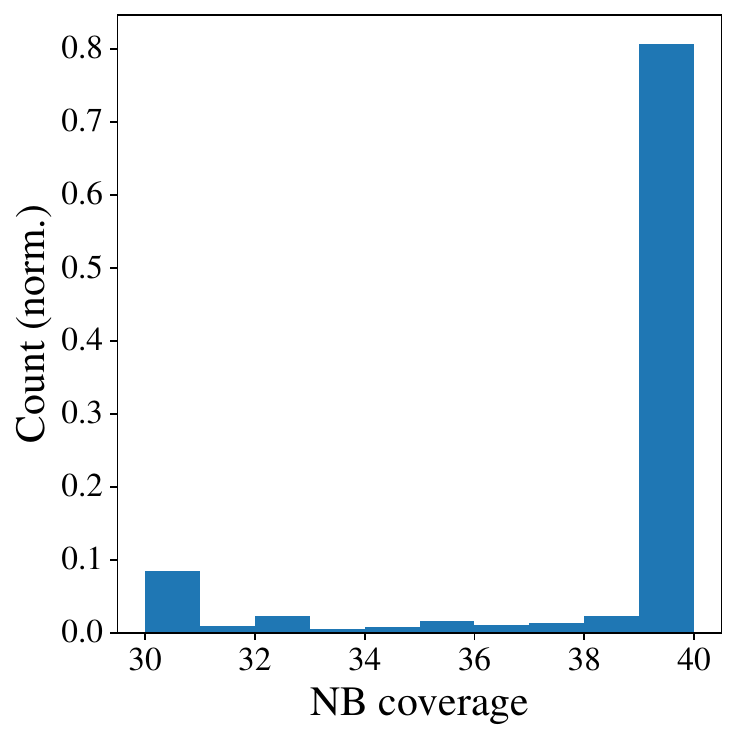}
   \includegraphics[width=0.4\textwidth]{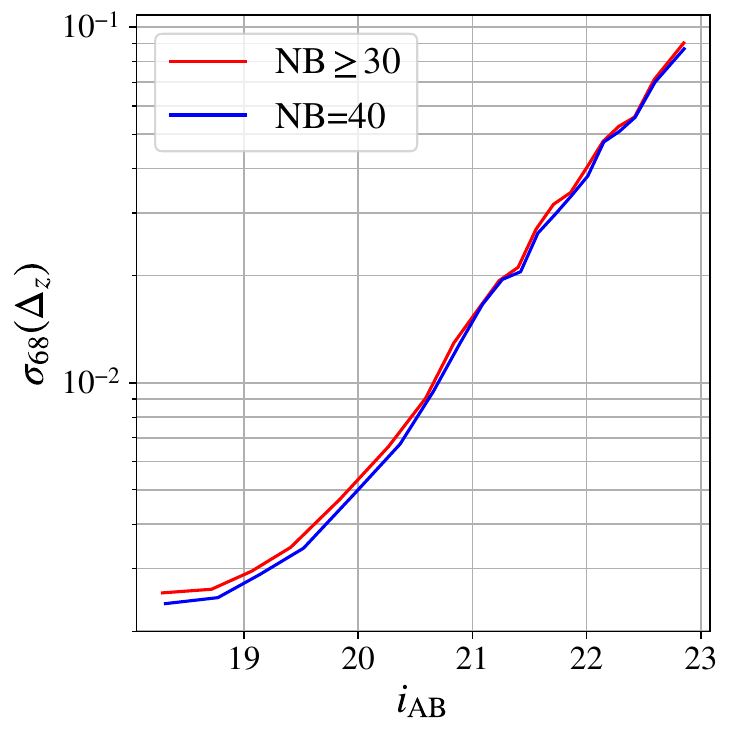}
    \caption{Left Panel: Normalised number of objects as a function of the NB coverage. Around 80\% of the objects with NB coverage equal or greater than 30 have measurements in all 40 NB. Right Panel: $\sigma_{68}$ as a function of $i_{\textrm{AB}}$ for objects in the PAUS wide fields with a NB coverage greater or equal than 30 NB (red) and for a NB coverage of 40 NB (blue). The degradation in the $\sigma_{68}$ is negligible in comparison with the number of objects recovered by including a more relaxed NB coverage condition.}
\label{fig:counts_sigma68_all_nrobs}
\end{figure*}

Left panel of Fig.~\ref{fig:counts_sigma68_all_nrobs} shows the number of objects as a function of the narrow band coverage. We decided to study the effect of a coverage of 30 NB in order not to lose much of the SED. Note that most of those objects have measurements in all 40 NB, while around $\sim$300000 objects lack some of the bands.

Right panel of Fig.~\ref{fig:counts_sigma68_all_nrobs} shows the $\sigma_{68}$ as a function of $i_{\textrm{AB}}$ for two cases: objects with NB coverage greater or equal than 30 NB (red) and objects with coverage in all bands (blue). The degradation in the $\sigma_{68}$ is almost negligible when reducing the coverage of narrow bands.

\section{Photo-\texorpdfstring{$z$}{z} catalogue}
\label{sec:Published catalogue}

We specify the column names and their description of the released catalogue in Table \ref{tab:photo-z_publish}.

\begin{table*}
\caption{Column name and its description for the published catalogue in CosmoHub.}
\begin{center}
\label{tab:photo-z_publish}
\begin{tabular}[c]{l l}
\hline
\hline
\textbf{Column name} & \textbf{Description} \\ \hline
ref$\_$id & PAUdm reference id (unique per PAUS wide field) \\ \hline
field & PAUS wide field \\ \hline
RA  & right ascension (deg) \\ \hline
DEC & declination (deg) \\ \hline
zb$\_$BCNZ & photometric redshift from BCNZ \\ \hline
zb$\_$BCNZw & weighted photometric redshift from BCNZ \\ \hline
odds & BCNZ ODDS quality parameter \\ \hline
chi2 & BCNZ minimum $\chi^{2}$ \\ \hline
nb$\_$bands & number of narrow bands \\ \hline
qz & BCNZ $Q_{z}$ quality parameter \\ \hline
mag$\_$i & $i_{\textrm{AB}}$ magnitude (not corrected for the difference between the CFHTLenS and KiDS magnitudes) \\ \hline
star$\_$flag & Only for objects from CFHTLenS: Star-galaxy separator (0 =galaxy, 1 =star) \\ \hline
mask$\_$cfhtlens & Only for objects from CFHTLenS: CFHTLenS mask value at the object’s position \\ \hline
sg$\_$flag & Only for objects from KiDS: Star/Gal Classifier (1 for galaxies) \\ \hline
sg2dphot & Only for objects from KiDS: 2DPhot Star/Galaxy classifier (1 for high confidence star) \\ \hline
mask$\_$kids & Only for objects from KiDS: 9-band mask information \\ \hline
\end{tabular}
\end{center}
\end{table*}


\bsp	
\label{lastpage}
\end{document}